\documentclass[superscriptaddress,twocolumn,showpacs,a4paper, amssymb,amsmath,nobibnotes,aps,prd, showkeys, nofootinbib,notitlepage]{revtex4-1}
\usepackage{verbatim}
\usepackage[T1]{fontenc}
\usepackage[utf8]{inputenc}
\usepackage[american]{babel}
\usepackage{epsfig}
\usepackage{booktabs}
\usepackage{multirow}
\usepackage{dcolumn}
\usepackage{amsmath}
\usepackage{mathtools}
\usepackage{amsfonts}
\usepackage{amssymb}
\usepackage{ulem}
\usepackage{epstopdf}
\usepackage{bm}
\usepackage{siunitx}
\usepackage{braket}
\usepackage{enumitem}
\usepackage{soul}
\usepackage[table]{xcolor}
\usepackage{color}
\usepackage{transparent}
\usepackage{pifont}
\usepackage{enumitem}

\definecolor{navyblue}{rgb}{0.0, 0.0, 0.5}
\definecolor{royalblue}{rgb}{0.25, 0.41, 0.88}
\definecolor{cadmiumgreen}{rgb}{0.0, 0.42, 0.24}
\definecolor{blue-violet}{rgb}{0.54, 0.17, 0.89}
\definecolor{darkviolet}{rgb}{0.58, 0.0, 0.83}
\definecolor{orange(colorwheel)}{rgb}{1.0, 0.5, 0.0}

\usepackage{hyperref}
\hypersetup{
    colorlinks=true, 
    linkcolor=royalblue, 
    citecolor=magenta}

\usepackage{booktabs}
\usepackage{multirow}
\usepackage{dcolumn}
\usepackage{colortbl}

\begin{document}

\title{A Model-Independent Test of Pre-Recombination New Physics: Measuring the Sound Horizon with Gravitational Wave Standard Sirens and the Baryon Acoustic Oscillation Angular Scale}

\author{William Giar\`e}
\email{w.giare@sheffield.ac.uk}
\affiliation{School of Mathematics and Statistics, University of Sheffield, Hounsfield Road, Sheffield S3 7RH, United Kingdom}

\author{Jonathan Betts}
\email{	jbetts3@sheffield.ac.uk}
\affiliation{School of Mathematics and Statistics, University of Sheffield, Hounsfield Road, Sheffield S3 7RH, United Kingdom}

\author{Carsten van de Bruck}
\email{c.vandebruck@sheffield.ac.uk}
\affiliation{School of Mathematics and Statistics, University of Sheffield, Hounsfield Road, Sheffield S3 7RH, United Kingdom}

\author{Eleonora Di Valentino}
\email{e.divalentino@sheffield.ac.uk}
\affiliation{School of Mathematics and Statistics, University of Sheffield, Hounsfield Road, Sheffield S3 7RH, United Kingdom}

\begin{abstract}
\noindent
In a broad class of cosmological models where spacetime is described by a pseudo-Riemannian manifold, photons propagate along null geodesics, and their number is conserved, upcoming Gravitational Wave (GW) observations can be combined with measurements of the Baryon Acoustic Oscillation (BAO) angular scale to provide model-independent estimates of the sound horizon at the baryon-drag epoch. By focusing on the accuracy expected from forthcoming surveys such as LISA GW standard sirens and DESI or Euclid angular BAO measurements, we forecast a relative precision of $\sigma_{r_{\rm d}} /r_{\rm d} \sim 1.5\%$ within the redshift range $z \lesssim 1$. This approach will offer a unique model-independent measure of a fundamental scale characterizing the early universe, which is competitive with model-dependent values inferred within specific theoretical frameworks. These measurements can serve as a consistency test for $\Lambda$CDM, potentially clarifying the nature of the Hubble tension and confirming or ruling out new physics prior to recombination with a statistical significance of $\sim 4\sigma$.

\end{abstract}


\maketitle


\textit{\textbf{Introduction --}} The disparity between the present-day expansion rate of the Universe -- quantified by the Hubble parameter $H_0$ -- as determined by the SH0ES collaboration using Type-Ia supernovae ($H_0=73\pm 1$ km/s/Mpc)~\cite{Riess:2021jrx,Murakami:2023xuy,Breuval:2024lsv}, and inferred by the Planck Collaboration from measurements of the Cosmic Microwave Background (CMB) temperature and polarization anisotropies angular power spectra, assuming a standard $\Lambda$CDM model of cosmology ($67.04\pm 0.5$ km/s/Mpc)~\cite{Planck:2018vyg}, has been regarded with significant attention by the cosmology community.

Barring any potential systematic origin of the discrepancy,\footnote{This scenario appears less likely following the extensive review conducted by the SH0ES collaboration, where several potential sources of systematics have been examined~\cite{Riess:2021jrx,Riess:2024ohe,Brout:2023wol}.} the Hubble tension might well stand as compelling evidence for the necessity of new physics beyond $\Lambda$CDM. While not aiming to offer a comprehensive overview of the debate that has heated up the community in the past few years,\footnote{We refer to Refs.~\cite{Agrawal:2019lmo,DiValentino:2021izs,Schoneberg:2021qvd,Abdalla:2022yfr,Poulin:2023lkg,Khalife:2023qbu,Giare:2024akf} for recent reviews on the Hubble tension problem, the related proposed solutions and their possible implications.} it is fair to say that attempts to tackle this issue have predominantly clustered around two distinct approaches: early-time and late-time solutions. Broadly speaking, the former category entails proposals suggesting new physics acting prior to recombination, while the latter comprises models that seek to modify the expansion history of the Universe after recombination.

This dichotomy in approach stems from the highly precisely determined angular scale of the acoustic peaks in the CMB spectra, $\theta_s$~\cite{Planck:2018vyg}, which sets the ratio between the sound horizon at recombination $r_s(z_*)$ and the angular diameter distance to the last scattering surface $D_A(z_*)$. Increasing the value of $H_0$ without disrupting the acoustic scale requires either a reduction in the value of the sound horizon -- a core tenet of early-time solutions -- or a distinct post-recombination expansion history of the Universe capable of compensating for a higher $H_0$ while preserving the angular diameter distance to the last scattering surface~\cite{Knox:2019rjx}.

Despite the apparent simplicity of the goals characterizing these two approaches, neither has proven effective in resolving the $H_0$-tension thus far, leaving the problem widely open. The primary challenge in the current landscape of solutions lies in the fact that both early and late-time observations tightly constrain new physics at their respective cosmic epochs, posing significant hurdles for model-building. Early-time solutions typically act near the last scattering surface, a cosmic epoch severely constrained by CMB measurements. Consequently, a common issue with early-time solutions is the need for a moderate level of fine-tuning to maintain a good fit to the CMB data, making it difficult to increase $H_0$ enough to match the SH0ES results~\cite{Jedamzik:2020zmd, Vagnozzi:2023nrq,DiValentino:2022fjm}.\footnote{In Ref.~\cite{Vagnozzi:2023nrq}, seven hints were proposed, suggesting that a compelling definitive solution might entail combining early and late-time new physics.}
On the other hand, late-time solutions are well constrained by observations of the local Universe, such as Baryon Acoustic Oscillations (BAO) and Supernovae (SN), which generally favor a $\Lambda$CDM-like cosmology at low redshifts~\cite{Efstathiou:2021ocp,Krishnan:2021dyb,Keeley:2022ojz,Gariazzo:2024sil}.\footnote{Recent BAO measurements released by the DESI collaboration~\cite{DESI:2024uvr,DESI:2024lzq,desicollaboration2024desi} seem to suggest dynamical dark energy, potentially reopening the avenue for new physical mechanisms at late times that could address the Hubble tension~\cite{Giare:2024smz}. See also Refs.~\cite{Wang:2024hks,Wang:2024dka,Wang:2024hwd,Cortes:2024lgw,Colgain:2024xqj,Yin:2024hba,Seto:2024cgo,Dinda:2024kjf} for discussions.} 

Given the difficulties in constructing successful models, it seems natural to step back and consider independent methods to test signals of new physics without relying on any particular framework. For late-time new physics, this is typically achieved through model-independent reconstructions of available datasets using Machine Learning (ML) techniques such as Artificial Neural Networks (ANN) or Gaussian Processes (GP). These approaches significantly limit theoretical assumptions in data analysis. Conversely, the state-of-the-art observational constraints on the early universe -- particularly at the time of recombination -- are largely dependent on the cosmological model assumed to analyze Planck CMB measurements. Therefore, a robust assessment of a fundamental physical quantity characterizing the early universe, not contingent on specific models, will certainly represent an important step forward for testing new physics before recombination and clarifying, once and for all, the intricate debate surrounding early and late-time solutions.

In this work, we demonstrate that geometric distance measurements in the local Universe can be used to independently measure the sound horizon. By using forecast measurements of Gravitational Waves (GWs) standard sirens from future surveys such as LISA~\cite{LISACosmologyWorkingGroup:2019mwx,LISACosmologyWorkingGroup:2022jok} for GWs, in conjunction with the angular scale of Baryon Acoustic Oscillations (BAO) anticipated by ongoing and forthcoming experiments such as DESI~\cite{DESI:2013agm,DESI:2016fyo} and Euclid~\cite{EUCLID:2011zbd,EuclidTheoryWorkingGroup:2012gxx,Amendola:2016saw}, we argue that it is possible to extrapolate the value of the sound horizon at the baryon drag epoch with a relative uncertainty of $\sim 1.5$\%, making no assumption about the cosmological model. Since the sound horizon encapsulates information about the Universe's expansion history from (soon after) the hot Big Bang singularity all the way up to recombination, this estimate can gauge early-time solutions, potentially confirming or ruling out new physics beyond $\Lambda$CDM with a statistical significance approaching 4 standard deviations. \bigskip

\textit{\textbf{Methodology -- }} To obtain a model-independent estimate of the sound horizon at the baryon-drag epoch,\footnote{It's important to highlight that BAO measurements are sensitive to the sound horizon evaluated at the baryon drag epoch, commonly denoted by $r_{\text{d}}$~\cite{Aubourg:2014yra}. Conversely, the scale pertinent to the acoustic peaks in the CMB is the sound horizon evaluated at recombination, typically denoted by $r_{s}(z_*)$~\cite{Hu:2001bc}. These two epochs are separated in redshift by $\Delta z = z_{\text{d}} - z_{*} \sim 30$.} we propose using the very simple relationship that links $r_{\rm d}$ and the angular scale of baryon acoustic oscillations:
\begin{equation}
\theta_{\rm BAO}(z) = \frac{r_{\rm d}}{(1+z) D_{\rm A}(z)} = \frac{(1+z) \, r_{\rm d}}{D_{\rm L}(z)},
\label{eq:BAO_DL}
\end{equation}
where, in the second equality, we have assumed the Distance Duality Relation (DDR), $D_{\rm L}(z)=(1+z)^2 D_{\rm A}(z)$, which connects the luminosity distance $D_{\rm L}(z)$ to the angular diameter distance $D_{\rm A}(z)$ at any redshift $z$.\footnote{This relation is quite general and is valid for any cosmological model where spacetime is described by a pseudo-Riemannian manifold, photons propagate along null geodesics, and their number is conserved over time. However, an important caveat is that in certain modified gravity theories, the distance inferred from standard sirens can differ from the (electromagnetic) luminosity distance. For instance, this discrepancy can arise in models featuring a running of an effective Planck mass, which rescales the luminosity due to modified friction in the GW propagation see, e.g., Refs.~\cite{Bellini:2014fua,Matos:2023jkn}. }

Starting from this relation, we can isolate $r_{\rm d}$ and express it in terms of $\theta_{\rm BAO}(z)$ and $D_{\rm L}(z)$. In principle, both of these quantities are directly measurable from current and future probes. So the next step involves identifying the most suitable datasets for estimating them.

\begin{itemize}

\item Acquiring a dataset capable of providing standard candles for measuring the luminosity distance, $D_{\rm L}(z)$, independently of any calibration methodologies or cosmological model assumptions, requires careful consideration. Although Type-Ia Supernovae serve as "standardizable" candles, they require calibration. Recent discussions have emphasized that the best achievable inference from supernovae data is the quantity $r_{\rm d} h$ (where $H_0=100 h$ km/s/Mpc)~\cite{Liu:2024lhu}, which does not provide sufficient information to address the nature of the Hubble tension and test new physics prior to recombination. Therefore, we suggest leveraging future gravitational wave observations as standard sirens to estimate $D_{\rm L}^{\rm GW}(z_{\rm GW})$. Planned experiments such as the Einstein Telescope (ET) and the Laser Interferometer Space Antenna (LISA) are expected to accurately measure the luminosity distance across a wide redshift range $z_{\rm GW} \in [z_{\rm GW}^{\rm min}, z_{\rm GW}^{\rm max}]$ that, depending on the specific experiment, can range from $z_{\rm GW}^{\rm min} \sim 0.1$ up to $z_{\rm GW}^{\rm max} \sim 8$.\footnote{ET will reach $z_{\text{GW}}^{\text{max}} \approx 2.5$, while LISA will cover redshifts up to $z_{\text{GW}}^{\text{max}} \approx 8$, see the \textit{supplementary material} for details.} For both ET and LISA, gravitational-wave standard sirens can achieve a few-percent precision on $D_{\rm L}(z)$ at $z \lesssim 0.5$. For LISA, favorable source orientations can yield relative uncertainties as low as $\sim$ 1–2\%, increasing up to $\gtrsim$ 10–30\% at higher redshifts or for less favorable orientations. For ET, relative uncertainties at low redshift are typically around 10\%, with few-percent precision achievable depending on redshift and inclination. However, they grow to $\sim$ 20–30\% at higher redshifts due to weak lensing and detector limitations. For more discussions we refer to the \textit{supplementary material}.

\item Measurements of the angular scales of BAOs have been released by several independent groups, extracting $\theta_{\text{BAO}}(z)$ from diverse catalogues, including those provided by the BOSS and eBOSS surveys~\cite{Nunes:2020hzy,deCarvalho:2021azj,Menote:2021jaq}. However, ongoing and forthcoming large-scale structure experiments, such as Euclid and DESI, are poised to significantly enhance the precision in estimating $\theta_{\text{BAO}}(z)$ across a redshift window $z_{\text{BAO}} \in [z_{\text{BAO}}^{\rm min}, z_{\text{BAO}}^{\rm max}]$, spanning from $z_{\text{BAO}}^{\text{min}} \sim 0.1$ to $z_{\text{BAO}}^{\text{max}} \sim 2$, contingent upon the specifics of the experiment.\footnote{DESI is covering redshifts $z_{\text{BAO}}^{\text{min}} \approx 0.15$ to $z_{\text{BAO}}^{\text{max}} \approx 1.85$, while Euclid is anticipated to collect data from $z_{\text{BAO}}^{\text{min}} \approx 0.65$ to $z_{\text{BAO}}^{\text{max}} \approx 2.05$.} To avoid mixing up current and forecast datasets, we focus on the anticipated improvements brought forth by DESI and Euclid-like experiments on $\theta_{\text{BAO}}(z)$, generating mock datasets for both surveys by following the methodology outlined in the \textit{supplementary material}. Note that, in both cases, the expected relative precision on $\theta_{\text{BAO}}(z)$ falls in the range of $1-4$\%, depending on the redshift bin.

\end{itemize}

Exploring various combinations of mock datasets from future BAO and GW surveys, we aim to forecast the precision achievable in determining $r_{\rm d}$ independently of any specific cosmological model. One significant challenge underlying the analysis arises from the fact that gravitational wave and BAO measurements will collect data at different redshifts, $z_{\text{GW}} \ne z_{\text{BAO}}$. To overcome this difficulty, we propose employing ML regression techniques, specifically ANN. As outlined in the \textit{supplementary material}, ANN can be trained to reconstruct the luminosity distance function, $D_{\text{L}}^{\text{GW}}(z)$, using gravitational wave observations $D_{\text{L}}^{\text{GW}}(z_{\text{GW}})$. Subsequently, these trained ANNs can extrapolate the luminosity distance to the same redshifts as BAO measurements: $D_{\text{L}}^{\text{GW}}(z_{\text{BAO}})$.\footnote{It is worth noting that GW data points are anticipated to outnumber BAO data points significantly. Therefore, we opt to reconstruct $D_{\text{L}}^{\text{GW}}(z)$ and estimate it at the same redshift as BAO measurements to enhance ML performance.}
In this way, we can obtain several independent estimates of $r_{\rm d}$ -- each corresponding to a BAO measurement:
\begin{equation}
r_{\rm d}= \frac{\theta_{\rm BAO}(z_{\rm BAO}) D_{\rm L}^{\rm GW}(z_{\rm BAO}) }{1+z_{\rm BAO}}.
\end{equation}
Based on the specifications of the various experiments under consideration, we identify a redshift range for each combination of datasets wherein both uncertainties and systematic errors are well under control. Sticking to these redshift ranges, we extract conservative yet informative estimates regarding the forecast precision of our measurement of the sound horizon, $\sigma_{r_{\text{d}}}$. For further technical details on this matter, we refer to the \textit{supplementary material}. \bigskip

\label{section:results}

\begin{table}[tpb!]
\begin{center}
\renewcommand{\arraystretch}{1.5}
\resizebox{ 0.95 \columnwidth}{!}{
\begin{tabular}{|c | c | c | c  |c |}
\hline 

\textbf{Redshift} & \textbf{DESI+ET} & \textbf{DESI+LISA} & \textbf{EUCLID+ET} & \textbf{EUCLID+LISA} \\

\hline \hline

$0.15$ & $4.3\%$ & $3.5\%$ & -- & --\\
$0.25$ & $7.5\%$ & $2.2\%$ & -- & --\\
$0.35$ & $9.7\%$ & $1.7\%$ & -- & --\\
$0.45$ & $11.5\%$ & $1.5\%$ & -- & --\\
$0.55$ & $13.1\%$ & $1.4\%$ & -- & --\\
$0.65$ & $14.8\%$ & $1.5\%$ & $14.2\%$ & $1.6\%$\\
$0.75$ & $16.2\%$ & $1.6\%$ & $15.5\%$ & $1.5\%$\\
$0.85$ & $17.5\%$ & $1.8\%$ & $16.9\%$ & $1.7\%$\\
$0.95$ & $18.6\%$ & $2.0\%$ & $18.2\%$ & $1.9\%$\\
$1.05$ & $19.6\%$ & $2.2\%$ & $19.5\%$ & $2.3\%$\\
$1.15$ & $20.4\%$ & $2.5\%$ & $20.6\%$ & $2.7\%$\\
$1.25$ & $21.2\%$ & $2.7\%$ & $21.5\%$ & $3.1\%$\\
$1.35$ & $21.8\%$ & $2.9\%$ & $22.2\%$ & $3.4\%$\\
$1.45$ & $22.4\%$ & $3.2\%$ & $22.6\%$ & $3.8\%$\\
$1.55$ & $22.9\%$ & $3.6\%$ & $22.8\%$ & $4.2\%$\\
$1.65$ & $23.5\%$ & $4.2\%$ & $23.0\%$ & $4.8\%$\\
$1.75$ & $24.0\%$ & $5.2\%$ & $23.2\%$ & $5.8\%$\\
$1.85$ & $24.6\%$ & $6.2\%$ & $23.3\%$ & $5.4\%$\\
$1.95$ & -- & -- & $23.5\%$ & $6.0\%$\\
$2.05$ & -- & -- & $24.12\%$ & $7.9\%$\\

\hline
\end{tabular}
}
\end{center}
\caption{\footnotesize Forecasted relative precision $\sigma_{r_{\rm d}}/r_{\rm d}$ on the sound horizon, derived from gravitational wave and baryon acoustic oscillation angular scale measurements by different combinations of surveys including ET and LISA for GWs, and DESI and Euclid for BAO.}
\label{tab.precision}
\end{table}

\textit{\textbf{Results -- }} In Tab.~\ref{tab.precision}, we summarize the relative precision we forecast on the sound horizon by combining measurements of gravitational waves and baryon acoustic oscillations from various surveys at different redshifts. To obtain these results, we assume conservative yet realistic forecasts for the number of gravitational-wave events with electromagnetic counterparts detectable by future observatories. For LISA, we follow the projections of Ref.~\cite{Tamanini:2016zlh} under the N2A5M5L6 configuration, which account for the expected capabilities of EM follow-up facilities. For the Einstein Telescope, we conservatively assume $\sim 10^3$ binary neutron star events with identified EM counterparts over several years of observation, consistent with cautious estimates in the literature~\cite{Branchesi:2023mws}. For a more detailed discussion of our setup, as well as an extensive set of consistency tests and analyses covering different configurations and event numbers, we refer to the \textit{Supplementary Material}.

At first glance, we note that when dealing with combinations involving ET, the uncertainties are quite substantial. For ET+Euclid, the percentage relative precision $\sigma_{r_{\rm d}}/r_{\rm d}$ consistently exceeds 10\%, reaching nearly 20\% at $z\sim 1$. These significant uncertainties arise from the fact that Euclid will be able to collect BAO measurements primarily at high redshifts $z\gtrsim 0.6$, where both the error bars of ET and the scatter of the mock data around the fiducial cosmology notably increase. This leads to broader uncertainty in the ML reconstruction of $D_{\rm L}^{\rm GW}(z)$ and, consequently, in the inferred value of $r_{\rm d}$.
On the other hand, combining DESI and ET allows for a significant reduction in uncertainties. This is because DESI is able to gather BAO measurements at lower redshifts where ET will observe more GW events with higher precision, resulting in a more accurate reconstruction of $D_{\rm L}^{\rm GW}(z)$. However, even with this overall improvement, the most optimistic estimate still yields $\sigma_{r_{\rm d}}/r_{\rm d} \gtrsim 4\%$, which remains comparable to the changes needed in $r_{\rm d}$ to resolve the Hubble tension.\footnote{We recall that to resolve the Hubble tension, early-time new physics should operate to decrease the value of the sound horizon by approximately $\Delta r_{\rm d}/r_{\rm d}\sim 5 - 6 \%$~\cite{Knox:2019rjx,Vagnozzi:2019ezj,Giare:2024smz}.} Moreover, uncertainties increase rapidly within the redshift range $z\sim0.25-0.55$, far exceeding $\sigma_{r_{\rm d}}/r_{\rm d} \gtrsim 10\%$. Therefore, the primary lesson we can glean is that the combinations of datasets involving ET are not ideal for achieving an informative model-independent measurement of $r_{\rm d}$. Such an estimate would carry uncertainties so significant that it would be practically unusable for assessing new physics at early times. 

Looking at the brighter side, a substantial improvement in constraining the value of the sound horizon occurs when we focus on LISA. Despite the fewer number of GW events expected from this survey, the error bars will be significantly reduced compared to ET, leading to a more precise reconstruction of the luminosity distance across the redshifts probed by DESI and Euclid. In this case, the constraining power achieved by combining LISA with either DESI or Euclid is similar: in both cases, we can select a notable redshift window ($z \lesssim 1$) where the uncertainties in the inferred value of the sound horizon remain $\sigma_{r_{\rm d}} /r_{\rm d}\lesssim 2\%$. In the most optimistic scenario, the uncertainties can be as small as $\sigma_{r_{\rm d}} /r_{\rm d}\lesssim 1.5\%$.

\begin{figure*}
    \centering
    \includegraphics[width=0.75\textwidth]{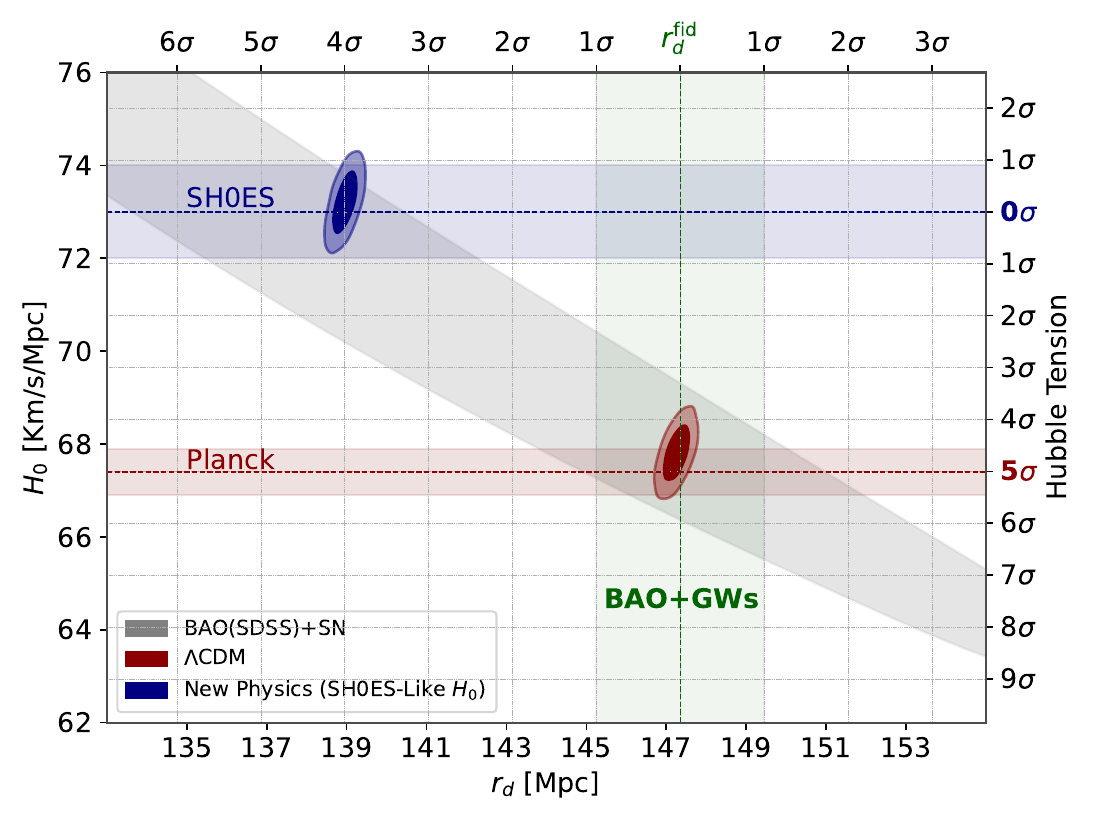}
    \caption{\footnotesize Illustrative plot in the $r_{\rm d}$ - $H_0$ plane of the consistency test proposed to assess the possibility of new physics prior to recombination for solving the Hubble constant tension. The red band represents the present value of $H_0$ measured by the Planck collaboration within a standard $\Lambda$CDM model of cosmology, whereas the 2D contours represent the marginalized 68\% and 95\% CL constraints obtained from the Planck-2018 data. The grey band represents the 95\% CL region of the plane identified by analyzing current BAO measurements from the SDSS collaboration and Type Ia supernovae from the Pantheon+ catalogue. The horizontal blue band represents the value of the Hubble constant measured by the SH0ES collaboration. In order to reconcile all the datasets, a potential model of early-time new physics should shift the $\Lambda$CDM red contours along the grey band until the grey band overlaps with the SH0ES result. This scenario is depicted by the 2D blue contours obtained under the assumption that the model of new physics does not increase uncertainties on parameters compared to $\Lambda$CDM. The green vertical band represents the model-independent value of the sound horizon we are able to extract from combinations of GW data from LISA and BAO measurements (either from DESI-like or Euclid-like experiments) assuming a fiducial $\Lambda$CDM baseline cosmology. As is clear from the top $x$-axis, this value would be able to confirm or rule out the possibility of new physics at about $4\sigma$.}
    \label{fig:1}
\end{figure*}

To demonstrate that a model-independent estimate of the sound horizon with a precision $\sigma_{r_{\rm d}} /r_{\rm d}\sim 1.5\%$ could decisively clarify the debate on early and late-time solutions while confirming or ruling out new physics beyond the standard cosmological model with high statistical significance, we propose the following conceptual exercise as a proof of concept. First, we suppose that the Hubble tension could indeed indicate the presence of new physics beyond the standard cosmological model. Second, we suppose the existence of an effective realization of early-time new physics -- though currently unknown -- that can reconcile the discrepancy. While remaining agnostic about the specific solution, we can outline the characteristics it must satisfy:

\begin{enumerate}

\item \textbf{Maintaining consistency with BAO and SN}: The solution must be consistent with all current cosmological datasets, including measurements of BAO and SN. As discussed in the literature~\cite{Knox:2019rjx}, the combination of SN and BAO data defines a band in the $r_{\rm d}$-$H_0$ plane. In Fig.~\ref{fig:1}, we show in gray the 95\% Confidence Level (CL) region of the parameter space defined by combining the SN gathered from the Pantheon Plus sample with BAO measurements released by the SDSS collaboration. Maintaining consistency with SN and BAO requires moving along this gray band.

\item \textbf{Consistency with SH0ES}: The solution must provide an $H_0$ value in agreement with the current local distance ladder estimate reported by the SH0ES collaboration, represented by the blue horizontal band in Fig.~\ref{fig:1}. Therefore, as we move along the gray band, we must intersect the blue band. A viable solution should lie at the intersection of the blue and gray bands in Fig.~\ref{fig:1}, implying a significant reduction in the sound horizon, as has been argued in the literature.

\item \textbf{Genuine Solution}: The solution must fit well with CMB data and genuinely resolve the tension by shifting the CMB contours for $\Lambda$CDM in the direction indicated by other datasets, ideally maintaining uncertainties in the Bayesian inference of cosmological parameters comparable to that achieved within the standard $\Lambda$CDM scenario.

\end{enumerate}

By imposing these three conditions, we can anticipate where the 2D probability contours that any hypothetical effective solution to the Hubble tension should place in the $r_{\rm d}$-$H_0$ plane. Our ideal scenario is depicted in Fig.~\ref{fig:1} with the blue 2D contours for the model that could resolve the Hubble tension. In the figure, we remain somewhat conservative and impose that the blue contours lie as close as possible to those obtained from the standard cosmological model -- that is, we consider the \textit{smallest} possible reduction in the value of the sound horizon that can fully satisfy these requirements.\footnote{Note that we could horizontally shift the 2D blue contours towards the left of the $x$-axis in Fig.~\ref{fig:1} while maintaining good agreement with all the three requirements we impose. This situation will imply a larger reduction in the sound horizon, increasing our ability to detect the change. In this sense, we work under reasonably conservative conditions.} Starting from this ideal scenario, we pose the following questions: supposing a theoretical physicist somewhere in the world, perhaps in the (near) future, realizes the solution represented by the blue contours in Fig.~\ref{fig:1}, \textit{(i)} is it possible to independently test this solution using the methodology introduced in this article, and \textit{(ii)} to what level of significance can we confirm or reject the hypothesis of new physics?

To address our questions, we can refer back to Fig.~\ref{fig:1}, where we illustrate the forecast obtained by assuming a standard $\Lambda$CDM cosmological model for the sound horizon, derived from a combination of data involving LISA, with a relative uncertainty of 1.5\%. As evident from the figure, a simple independent estimate of the sound horizon would represent a remarkable consistency test for $\Lambda$CDM, dismissing the hypothesis of new physics at a statistical significance exceeding 4 standard deviations (refer to the $x$-axis at the top of the figure). At this point, only two possibilities would remain on the table: the first would be to assume that the Hubble tension is due to systematics. This possibility, in turn, can be tested with gravitational waves as standard sirens, which will precisely measure $H_0$~\cite{Chen:2017rfc,Palmese:2019ehe}, confirming or rejecting the measurement obtained from SH0ES. If the Hubble tension were confirmed, then this would necessarily imply the need to resort to a different physical mechanism than the one proposed by early-time solutions. In our ideal scenario to resolve the problem, we will need to shift the red contours in Fig.~\ref{fig:1} vertically and force the cosmological model to lie at the intersection between the green and blue horizontal bands. This possibility would be in strong tension with the gray contours derived from current SN and BAO measurements; thus, it appears less plausible.

On the flip side, it is worth noting that this argument works in reverse, too. If the hypothetical theoretical physicist happens to have stumbled upon the correct solution, our test would yield a model-independent measurement of the sound horizon consistent with that predicted by the model of new physics, thereby ruling out the value inferred within the standard cosmological framework at 4 standard deviations.

In closing, we would like to stress two key remarks. First, as shown in the \textit{Supplementary Material}, although the relative precision on $r_d$ clearly depends on the number of gravitational-wave detections, our methodology remains robust even under pessimistic assumptions: with as few as 7-10 low-redshift ($z < 0.5$) detections from LISA, the sound horizon can be measured at the 1.4-2.6\% level, while with only 4-6 detections, we achieve 2-3.5\% precision, demonstrating the resilience of the approach in low-statistics regimes. Second -- and most importantly --  the significance of our test lies in the fact that gravitational wave standard sirens will be able to extract model-independent measurements of the sound horizon with a precision competitive with current model-dependent estimates from early Universe probes. This will allow us to test whether the value of the sound horizon is consistent with those predicted by the standard cosmological model, providing an important tool to shed light on the nature of the Hubble tension. Among other things, this will clarify once and for all the debate on early vs late-time solutions and set clear model-building guidelines for assessing possible new physics beyond $\Lambda$CDM. \bigskip


\textit{\textbf{Conclusion -- }} In this work, we propose a method for measuring the sound horizon at the baryon drag epoch, $r_{\rm d}$, independently of the cosmological model. Our methodology is based on the relationship linking the angular scale of baryon acoustic oscillations and the angular diameter distance. Assuming that spacetime is described by a pseudo-Riemannian manifold, photons propagate along null geodesics, and their number is conserved over time, we can use the distance duality relation to connect $\theta_{\rm BAO}(z)$ and $D_{\rm L}(z)$ by means of Eq.~\eqref{eq:BAO_DL}. Starting from this relation, we propose using future gravitational waves as standard sirens to gather precise measurements of the luminosity distance in the Universe, which are free from calibration methods compared to current "standardizable" candles such as Type-Ia supernovae.

We argue that employing machine learning techniques, particularly artificial neural network linear regression, one can combine future gravitational wave observations from LISA with angular BAO measurements, either from DESI-like or Euclid-like surveys, enabling us to obtain model-independent estimates of the sound horizon. In the redshift range $z \lesssim 1$, we forecast that the relative precision of these estimates will be as small as $\sigma_{r_{\rm d}} /r_{\rm d} \sim 1.5\%$. This would offer a \textit{unique} model-independent measure of a fundamental scale characterizing the early universe, with a precision competitive to values inferred within specific theoretical frameworks, including the standard cosmological model. Such measurements have several significant implications:

\begin{enumerate}

\item[\textit{(i)}] \textbf{Providing a consistency test for the standard cosmological model:} We can test the value of $r_{\rm d}$ predicted by the baseline cosmological model with high statistical significance, either confirming its predictions or providing hints of new physics beyond $\Lambda$CDM.

\item[\textit{(ii)}]  \textbf{Shedding light on the nature of the Hubble tension:} If the tension is confirmed by gravitational wave standard sirens, our test could provide crucial information about its nature. In particular, we will be able to conclusively assess if the Hubble tension requires new physics acting prior to recombination to reduce the value of the sound horizon. As shown in this work, we can draw some general model-agnostic guidelines from current data that give solid grounds to believe that a compelling early-time solution would require a reduction of about 5-6\% in the value of the sound horizon. Such a hypothesis can be confirmed or ruled out at least at $4\sigma$ by our test, see also Fig.~\ref{fig:1}.

\item[\textit{(iii)}]  \textbf{Providing a definitive answer about the debate surrounding early vs late-time solutions of the Hubble tension:} Should the Hubble tension be confirmed by gravitational wave standard sirens, leading to a necessary paradigm shift in cosmological modeling, our test can provide definitive guidelines. It will decisively address the ongoing debate between early and late-time solutions. On one hand, the detection of new physics from the early universe with high statistical significance could bolster early-time solutions, challenging the late-time community. Conversely, if our test confirms the sound horizon value predicted by the $\Lambda$CDM model, it would undermine early-time solutions, thus favoring alternative late-time proposals.

\end{enumerate}

\textit{\textbf{Acknowledgments -- }} We thank Elsa M. Teixeira,  Richard Daniel and Isabela Matos for the help and useful discussions. WG and CvdB are supported by the Lancaster–Sheffield Consortium for Fundamental Physics under STFC grant: ST/X000621/1.
EDV is supported by a Royal Society Dorothy Hodgkin Research Fellowship. This article is based upon work from COST Action CA21136 Addressing observational tensions in cosmology with systematics and fundamental physics (CosmoVerse) supported by COST (European Cooperation in Science and Technology). We acknowledge IT Services at The University of Sheffield for the provision of services for High Performance Computing.

\bibliographystyle{apsrev4-1}
\bibliography{bib}


\widetext
\clearpage

\begin{center}
\huge{\textsc{Supplementary Material}}  
\end{center}

\vspace{1.5 cm}

\section{Forecasts of the BAO Angular Scale} 
\label{section:appendix-BAO}
 
To generate forecasted measurements, both for the angular BAO and gravitational waves (detailed in the next section), we adopt a fiducial $\Lambda$CDM cosmological model. We set the values of cosmological parameters to their best-fit values from the Planck 2018 analysis \cite{Planck2018}, as summarized in Table~\ref{tab.Priors}. Fixing these parameter values, we derive other relevant quantities and compute the theoretical model using the Boltzmann solver code \texttt{CLASS} \cite{class1, class2, class3}.

\begin{table}[htb!]
\begin{center}
\renewcommand{\arraystretch}{1.5}
\begin{tabular}{l@{\hspace{0. cm}}@{\hspace{1.5 cm}} c}
    \hline
    \textbf{Parameter}    & \textbf{Fiducial Value} \\
    \hline\hline
    $\Omega_{\rm b} h^2$         & $0.022383$ \\
    $\Omega_{\rm c} h^2$         & $0.12011$ \\
    $\log(10^{10}A_{\rm S})$     & $3.0448$ \\
    $n_{\rm s}$                  & $0.96605$ \\
    $100\,\theta_{\rm {MC}}$     & $1.040909$ \\
    $\tau$                       & $0.0543$ \\
    $\sum m_{\nu}$ [eV]                & $0.06$ \\
    \hline\hline
\end{tabular}
\caption{Fiducial Value of Cosmological Parameters.}
\label{tab.Priors}
\end{center}
\end{table}

To forecast the angular BAO scale, $\theta_{\rm BAO}(z)$, we follow established methods previously employed to test and constrain extensions of $\Lambda$CDM~\cite{DiValentino:2018jbh,Giare:2021cqr}. Our focus in this study is on forecasting for the upcoming Dark Energy Spectroscopic Instrument (DESI)~\cite{DESI:2013agm,DESI:2016fyo} and Euclid~\cite{EUCLID:2011zbd,EuclidTheoryWorkingGroup:2012gxx,Amendola:2016saw} redshift surveys. Starting with the fiducial $\Lambda$CDM cosmology, we simulate measurements of the angular diameter distance at the redshifts where DESI-like and Euclid-like experiments are expected to collect data. We assume a realistic relative precision, $\sigma_{D_{\rm A}}(z) / D_{\rm A}(z)$, which varies with redshift and survey. Information on the redshifts and relative precision is obtained from Table V and Table VI of Ref.~\cite{DESI-EUCLID:values} for DESI-like and Euclid-like surveys, respectively.

We introduce Gaussian noise in the mock measurements of $D_{\rm A}(z)$ to simulate the final observations. Various statistical metrics are considered to determine the appropriate level of scatter around the fiducial value. Ultimately, we adopt the following criterion: for each BAO measurement, we ensure that the difference between the fiducial value and the forecast data point, normalized by the observational uncertainties, is smaller than 1:
\begin{equation}
\frac{|D_{\rm A}(z)-D_{\rm A}^{\rm{fid}}(z)|}{\sigma_{D_{\rm A}(z)}}\lesssim 1.
\end{equation}
This is already mostly true for current SDSS BAO measurements (see the bottom panel in Fig. 6 of Ref.~\cite{Escamilla_2024}) and the first data release of DESI (see the bottom panel in Fig. 2 of Ref.~\cite{Giare:2024smz}).\footnote{We test that this choice does not significantly affect the conclusions reached in this paper, as well as that we can recover the fiducial cosmological model when performing Markov Chain Monte Carlo (MCMC) analyses for these data, see Sec.~\ref{sec:tests}.}

\begin{figure}[htb!]
    \centering
    \includegraphics[width=0.6\columnwidth]{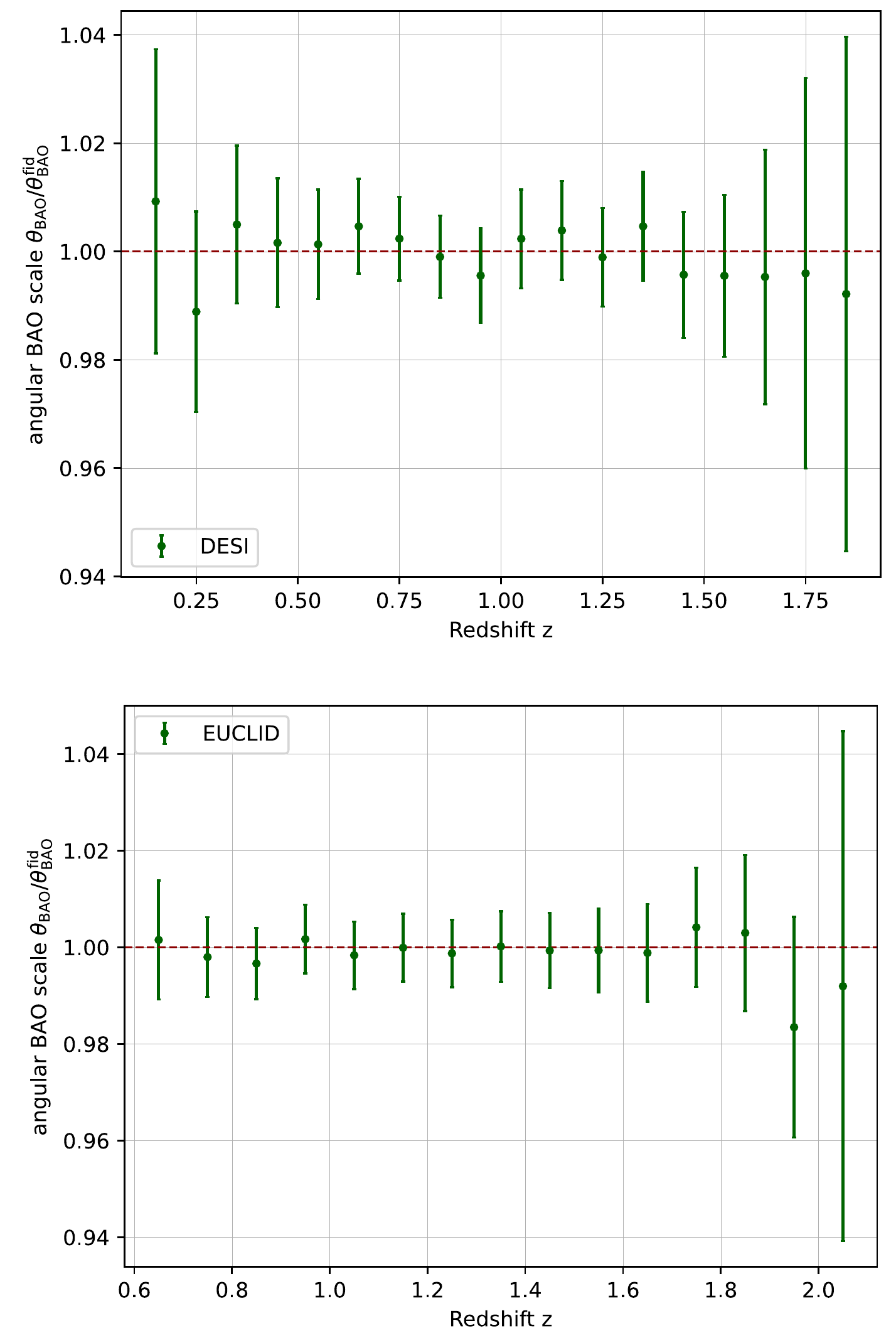}
    \caption{Example forecast for DESI (top panel) and Euclid (bottom panel) observations. The plots show the forecasted values of the BAO angular scale $\theta_{\rm BAO}(z) / \theta^{\rm fid}_{\rm BAO}(z)$, where $\theta^{\rm fid}_{\rm BAO}(z)$ is the angular BAO scale predicted by the fiducial $\Lambda$CDM cosmology. The error bars indicate the uncertainties in these forecasted values. The forecasted data points are generated based on the expected performance of the DESI and Euclid surveys.}
    \label{fig:BAOs}
\end{figure}

The BAO angular scale $\theta_{\rm BAO}(z)$, and the corresponding error $\sigma_{\theta}(z)$ can then be computed from $D_{\rm A}(z)$ and $\sigma_{D_{\rm A}}$ as
\begin{equation}
    \theta_{\rm BAO}(z) = \frac{\kappa r_{\rm d}^{\rm fid}}{(1+z) D_{\rm A} (z)},  \quad \sigma_{\theta}(z) = \frac{\kappa r_{\rm d}^{\rm fid} \sigma_{D_{\rm A}}}{(1+z) D_{\rm A}(z)^2}
\end{equation}
with $\kappa = 180/\pi$ and $r_{\rm d}^{\rm fid}$ being the sound horizon corresponding to the fiducial $\Lambda$CDM cosmology ($r_{\rm d}^{\rm fid} \sim 147$ Mpc). We show in Fig.~\ref{fig:BAOs} the forecasted values for $\theta_{\rm BAO}(z)$ obtained following this methodology.

\section{Forecasts of Gravitational Waves as Standard Sirens} 
\label{section:appendix-GW}

To reconstruct the redshift-luminosity distance relation $D_L(z)$ from gravitational waves used as standard sirens through an artificial neural network, it is essential to have training data consisting of a series of $D_L(z)$ measurements along with their associated errors $\sigma_{D_L}$. In this work, we generate mock data to simulate gravitational wave measurements that are expected from future experiments such as the Einstein Telescope (ET) and the Laser Interferometer Space Antenna (LISA). For producing these data, we strictly follow the methodology proposed in Ref.~\cite{Teixeira:2023zjt}, which we summarize step-by-step below.

\subsection{Merger Event Distribution}
Given the $\Lambda$CDM background cosmology, a redshift distribution of gravitational-wave-generating merger events is generated for each of the ET and LISA experiments based on the events those experiments are designed to detect. For ET, these events are the mergers of binary neutron stars (BNS) and black hole neutron star binaries (BHNS), and follow the distribution $P$ defined as
\begin{equation}
    P \propto \frac{4\pi d_c(z)R(z)}{(1+z)H(z)},
\end{equation}
where $d_c(z)$ and $H(z)$ represent the comoving distance and Hubble parameter, respectively. These quantities, for our model of cosmology, are obtained via \texttt{CLASS} at various redshifts. $R(z)$ is known as the merger rate for BNS and BHNS systems, defined to a linear approximation in Ref.~\cite{ET_merger_rate} as
\begin{equation}
     R(z) = \begin{cases} 
          1+2z & \text{for } z < 1, \\
          \frac{3}{4}(5-z) & \text{for } 1 \leq z < 5, \\
          0 & \text{otherwise}.
       \end{cases}
\end{equation}
In this work, we consider simulated ET events in the range $0 \leq z \leq 2.5$. The LISA experiment targets lower frequencies than ET and therefore focuses on higher-mass binary massive black hole (BMBH) mergers in the range $0 \leq z \leq 8$. These events were generated based on the L6A2M5N2 mission distribution \cite{Caprini_2016}. 

For GW events to serve as standard sirens, the identification of an electromagnetic (EM) counterpart is crucial, as it enables precise localization of the source. In our simulations for the upcoming LISA and ET datasets, we assume the existence of such counterparts, following the justification in \cite{ET_merger_rate, Caprini_2016}, which also informs the GW event distributions used in this work. By the time LISA and ET are operational, relevant EM data will be available from major observational facilities, including the Large Synoptic Survey Telescope (LSST)\footnote{\url{www.lsst.org}}, the Square Kilometer Array (SKA)\footnote{\url{www.skao.int}}, and the Extremely Large Telescope (ELT)\footnote{\url{www.eso.org/sci/facilities/eelt}}. To ensure robust forecasts, we adhere strictly to the specifications outlined in the LISA~\cite{Babak:2017tow} and ET~\cite{ET:2019dnz} White Papers, which provide estimates for EM counterpart detection. Specifically:

\begin{itemize}

\item For LISA, our modeling is based on the L6A2M5N2 mission configuration described in Ref.~\cite{Caprini:2016qxs}, with detectable EM counterparts estimated in line with forecasts such as those in Ref.~\cite{Tamanini:2016zlh}. In particular, the N2A5M5L6 setup yields a comparable number of well-localized events under similarly conservative assumptions. These estimates incorporate the expected capabilities of EM follow-up facilities at the time of observation (see Section 5.3 and Tables 9–10 of Ref.~\cite{Tamanini:2016zlh}) and are consistent with those adopted in our analysis. Specifically, the sky localization of sources is typically better than $10 , \text{deg}^2$, aligning well with the field of view of upcoming optical and radio surveys such as LSST and SKA. Nearby extreme-mass-ratio inspirals (EMRIs) at $z<0.1$ are often localized to within $1 , \text{deg}^2$, facilitating host galaxy identification and independent redshift measurements. As we will prove in Sec.\ref{sec.GWs_tests}, these low-redshift GW detections are crucial for our methodology to succeed. Therefore, it is particularly important to adopt a conservative approach in our forecasts. In this regard, we assume only \textit{7-10} detectable EMRIs at $z < 0.5$ (see the bottom panel of Fig.\ref{fig:GW_recon}), significantly below the $\sim 20$ events forecast in previous studies. For context, in Ref.~\cite{MacLeod:2007jd}, MacLeod $\&$ Hogan demonstrated that detecting $\sim 20$ EMRI events at $z < 0.5$ with LISA would enable a determination of $H_0$ at the $< 1\%$ level via statistical redshifts from galaxy surveys. The LISA White Paper predicts that nearly all models forecast more than 20 such events at $z < 0.5$. Our assumption of \textit{7-10} events is, therefore, somewhat conservative compared to what is typically adopted in the literature.

\item Similarly, for ET, binary neutron star (BNS) mergers will be detectable up to $z \sim 2-3$, with annual detection rates reaching $\sim 7 \times 10^4$ for BNS coalescences and up to $10^6$ for binary black holes (BBHs). The number of events with an observed EM counterpart will depend on the available EM facilities at that time, but estimates suggest that over several years, one could collect $\gtrsim \mathcal{O}(10^3)$ BNS GW events with identified EM counterparts. Again, we adopt a conservative threshold of $\sim 10^3$ detected events, which is more cautious than the assumptions commonly used in the literature.

\end{itemize}

In Sec.~\ref{sec.GWs_tests}, we will assess the constraining power of our GW forecasts and compare them with the results typically documented in the literature.

\subsection{Simulating $D_L(z)$ Measurements and $\sigma_{D_L}$ Errors}
Once the standard siren catalogues $D_L(z)$ have been generated from their respective redshift distributions, the events that could plausibly be measured by the LISA and ET experiments must be identified, and the error associated with those measurements $\sigma_{D_L}$ computed. This is done based on the signal-to-noise ratio (SNR) for a given event. SNR is defined as
\begin{equation}
     \text{SNR} = 4\int_{f_{\text{min}}}^{f_{\text{max}}} \frac{|\mathcal{H}|}{S_{h}} \, df,
\end{equation}
where $\mathcal{H}$ is the Fourier transform of the strain on the LISA/ET interferometer and $S_{h}$ is the noise power spectral density. $S_{h}$ weights the SNR depending on the specific properties of the instruments used in the ET and LISA detectors (see equations 15 and 16 of Ref.~\cite{Teixeira:2023zjt}). Siren signals in the GW catalogue are discarded as unlikely to be detected by either LISA or ET if $\text{SNR} < 8$.

After SNR slicing of the siren catalogue, the error $\sigma_{D_L}$ is computed using the sum of error contributions
\begin{equation}
     \sigma_{D_L} = \sqrt{(\sigma^{\rm inst}_{D_L})^2+(\sigma^{\rm len}_{D_L})^2+(\sigma^{\rm pec}_{D_L})^2},
\end{equation}
where $\sigma^{\rm inst}_{D_L} \approx 2D_L/\rho$ is the instrumental error in the luminosity distance measurement, $\sigma^{\rm len}_{D_L}$ is the error due to gravitational lensing 
\begin{equation}
     \sigma^{\rm len}_{D_L} = \frac{D_L}{2} \times 0.066 \left[4(1-(1+z)^{1/4})\right]^{1.8}
\end{equation}
and $\sigma^{\rm pec}_{D_L}$ is a factor unique to the orbital LISA experiment ($\sigma^{\rm pec}_{D_L} = 0$ for ET) derived from the peculiar velocities of GW sources relative to the Hubble flow
\begin{equation}
     \sigma^{\rm pec}_{D_L} = D_L \frac{\sqrt{\langle v^2 \rangle}}{c} \left[1 + \frac{c(1+z)}{H D_L}\right],
\end{equation}
with estimate $\sqrt{\langle v^2 \rangle} = 500\ \text{km s}^{-1}$. 

\begin{figure}[htb!]
    \centering
    \includegraphics[width=0.6\columnwidth]{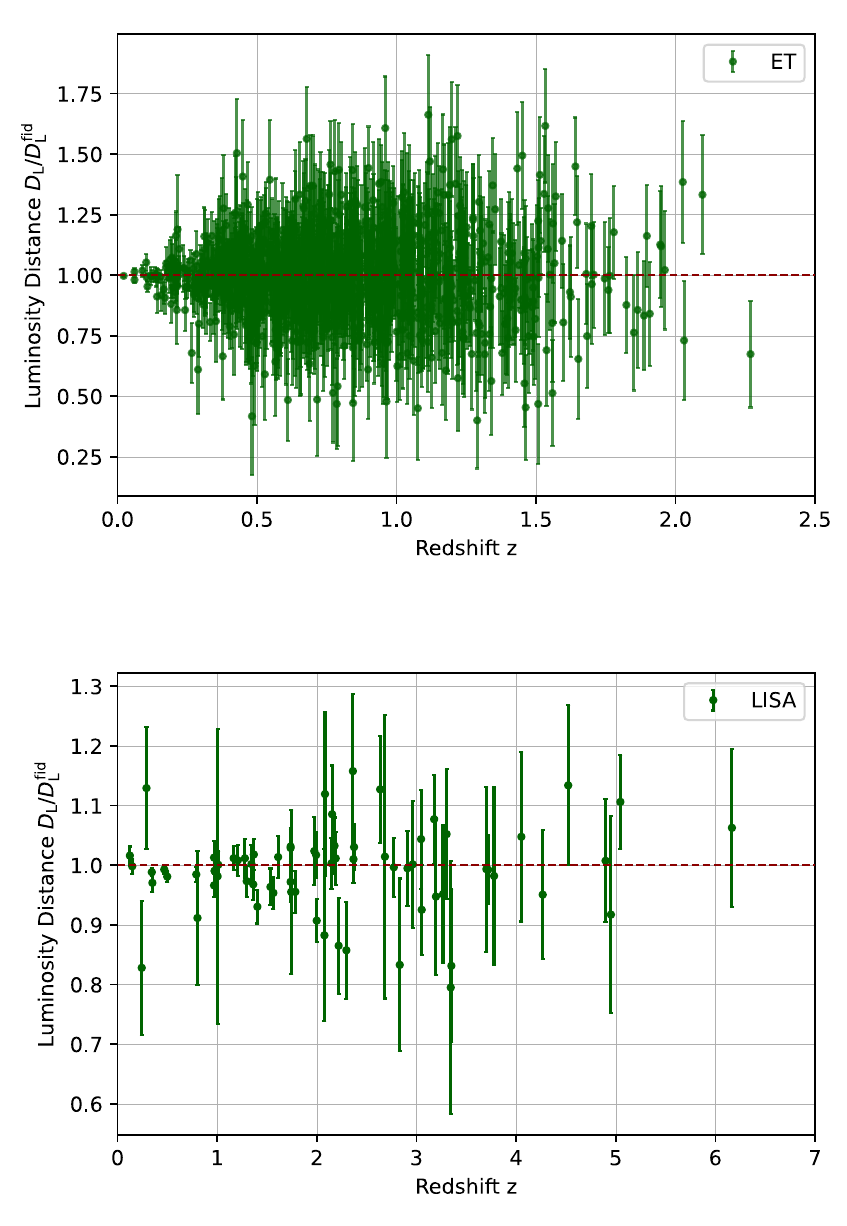}
    \caption{Example forecast for ET (top panel) and LISA (bottom panel) observations. The plots show the forecasted values of the luminosity distance $D_L(z) / D_L^{\rm fid}(z)$, where $D_L(z)^{\rm fid}$ is the luminosity distance predicted by the fiducial $\Lambda$CDM cosmology. The error bars indicate the uncertainties in these forecasted values. The forecasted data points are generated based on the expected performance of the ET and LISA surveys.}
    \label{fig:GWs}
\end{figure}

In Fig.~\ref{fig:GWs} we show the forecasted values for $D_L(z)$ obtained following this methodology. For LISA, favorable source orientations and low redshifts $z \lesssim 0.5$ can lead to relative precisions as tight as $\sim 1-2 \%$. For supermassive black hole binaries in the range $z \sim 1-4$, relative uncertainties of $\sim$1-10\% are expected, assuming accurate sky localization and waveform modeling. In particularly high signal-to-noise events with identified host galaxies, precision may improve to a few percent. At higher redshifts, however, uncertainties can degrade to $\gtrsim 10-30$\%. For ET, binary neutron star (BNS) mergers with electromagnetic counterparts can yield relative uncertainties on $D_{\rm L}(z)$ typically in the range of 1-10\%, depending on redshift and inclination. For favorable orientations and at low redshifts $z \lesssim 0.5$, measurements of $D_{\rm L}(z)$ with precision at the few-percent level are possible. However, the vast majority of events are expected to yield uncertainties around $\sim$10\%. At higher redshifts $z \sim 2$, uncertainties can increase up to $\sim 20-30$\%, due in part to weak lensing and detector limitations. 

\newpage 
\subsection{ANN Reconstruction of $D_{\rm L}(z)$} 
\label{section:appendix-reconstruction}

In order to combine $D_{\rm L}(z)$ from GW sirens with BAO observations at arbitrary redshifts to compute $r_{\rm d}$, a continuous $D_{\rm L}(z)$ function must be reconstructed from the $D_{\rm L}(z)$ and $\sigma_{D_L}$ datasets. This work uses the \texttt{refann}\footnote{\url{https://github.com/Guo-Jian-Wang/refann}} code \cite{Wang_2020} to construct an artificial neural network (ANN)\footnote{ANNs are known to be equivalent to Gaussian Processes in certain limits \cite{2017arXiv171100165L} and are designed for reconstructing functions from precisely this type of data.} that describes a continuous $D_{\rm L}(z)$ function for both the LISA and ET experiments. A common justification for choosing an ANN architecture for this type of function reconstruction problem is the set of universal approximation theorems for bounded and unbounded network width and depth \cite{CYBENKO1989, HORNIK1989359, HORNIK1991251, LESHNO1993861, GULIYEV2018296}. This reasoning is quite broad and only states the existence of such an optimal approximation, not that it is computationally tractable. Other methods such as splines and regression techniques are conventionally used in the literature to study $D_{\rm L}(z)$ within the limit of $z<3$ \cite{mangiagli2024massiveblackholebinaries}. However, ANNs require no assumptions regarding the underlying functional form of the data, are known to be equivalent to Gaussian Processes in certain limits \cite{2017arXiv171100165L}, are more extensible and are designed for reconstructing functions from precisely this type of data.

\begin{figure*}[htbp]
    \centering
    \includegraphics[width=0.8\textwidth]{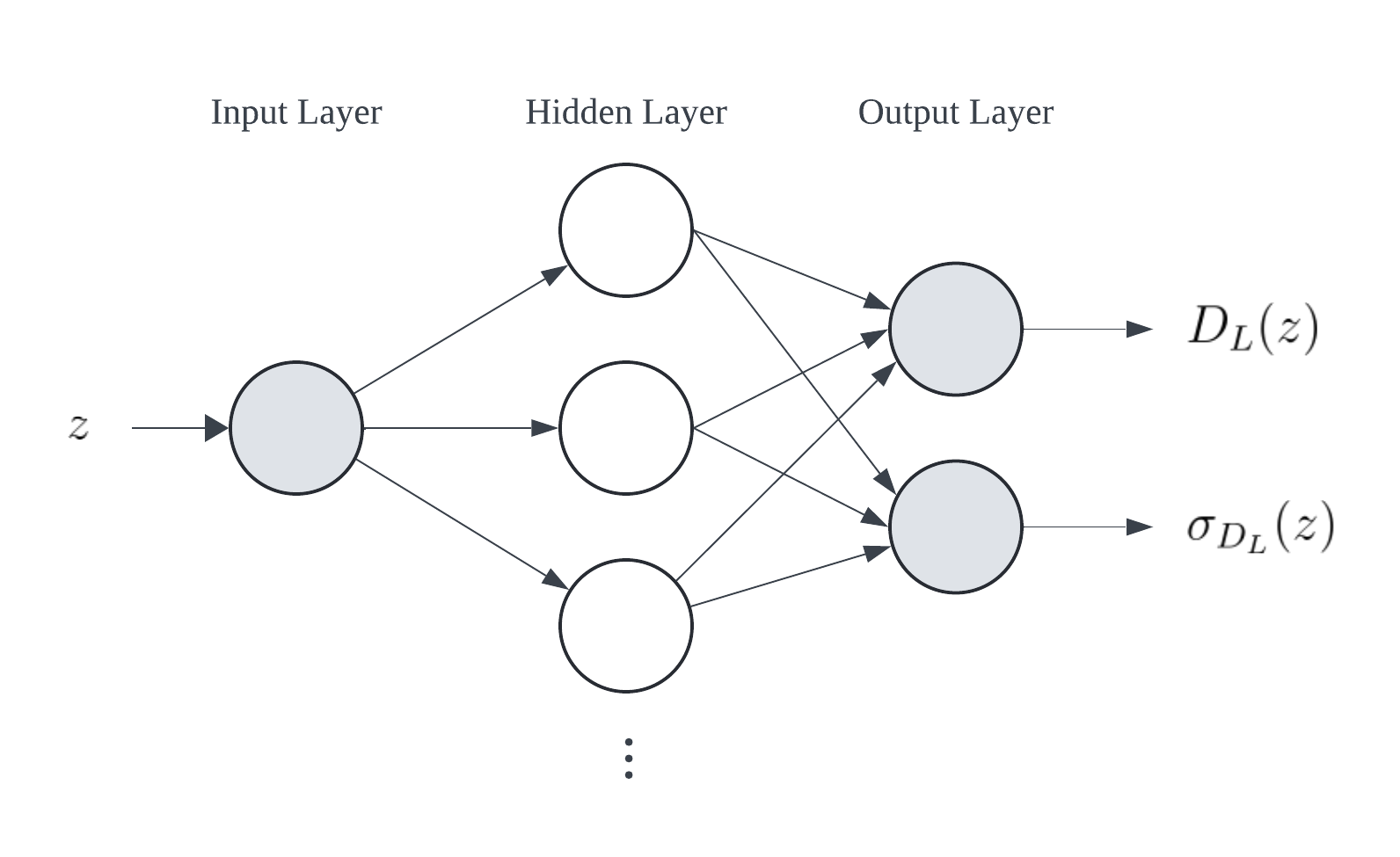}
    \caption{Architecture of the single-hidden-layer ANN used to reconstruct $D_{\rm L}$ and $\sigma_{D_{\rm L}}$ at input redshift $z$.}
    \label{fig:ann_arc}
\end{figure*}

Fig.~\ref{fig:ann_arc} shows the architecture of the ANN used in this work. In general, ANNs consist of an input layer, one or more hidden layers and an output layer. Each layer accepts a vector from the previous layer and applies a weighted linear transformation followed by a nonlinear activation function, before passing the result onto the next layer. It is these weights at each layer of the network that are optimised during supervised training based on a chosen loss function against a labelled target output. This process is formalized in vector notation as
\begin{equation}
    \mathbf{z}_{i+1} = \mathbf{x}_{i}\mathbf{W}_{i+1} + \mathbf{b}_{i+1},
    \label{eq:ann1}
\end{equation}
\begin{equation}
    \mathbf{x}_{i+1} = f(\mathbf{z}_{i+1}),
    \label{eq:ann2}
\end{equation}
where $\mathbf{x}_{i}$ is the $i$-th layer input vector, $f$ is the nonlinear activation function and $\mathbf{W}_{i+1}$ and $\mathbf{b}_{i+1}$ are the linear weights and biases to be optimised. 
The ANN in this work uses one hidden layer with an elementwise Exponential Linear Unit \cite{clevert2016fast} activation function
\begin{equation}
     f(x) = \begin{cases} 
          x & \text{for } x > 0 \\
          e^x - 1 & \text{for } x \leq 0.
          \end{cases}
\end{equation}

In the practical implementation of such a network, these vector processes are performed in batches of inputs $X \in \mathbb{R}^{l \times n}$ where $n$ is the dimensionality of each input and $l$ is the number of inputs in the batch. Equations \eqref{eq:ann1} and \eqref{eq:ann2} then become the matrix process
\begin{equation}
    Z_{i+1} = X_{i}W_{i+1} + B_{i+1},
    \label{eq:ann3}
\end{equation}
\begin{equation}
    X_{i+1} = f(Z_{i+1}),
    \label{eq:ann4}
\end{equation}
such that the ANN can now be understood in terms of the function $f_{W,B}$ on input $X$. In a supervised learning setting, each input in $X$ is labelled with its corresponding ground-truth output $Y_{\rm truth} \in \mathbb{R}^{l \times p}$ where $p$ is the dimensionality of the output. Training the model then becomes minimizing the difference between predicted outputs $Y_{n, \text{pred}} = f_{W,B}(X_n)$ and $Y_{n, \text{truth}}$ for the $n$-th input in $X$, by optimizing the parameter sets $W$ and $B$. This is done using the mean-absolute-error function
 \begin{equation}
    \mathcal{L} = \frac{1}{lp}||Y_{\text{pred}} - Y_{\text{truth}}||,
    \label{eq:mae}
\end{equation}
root-mean-squared error was also tested, and a negligible difference in model performance was found. Equation \eqref{eq:mae} is then used for optimization of weights $W$ and biases $B$ by iterative algorithmic back-propagation. 
Back-propagation involves computing the following derivatives of $\mathcal{L}$ \cite{backprop}:
\begin{equation}
    \frac{\partial\mathcal{L}}{\partial Z_{i+1}} = f'(Z_{i+1})\frac{\partial\mathcal{L}}{\partial X_{i+1}},
    \label{eq:bp1}
\end{equation}
\begin{equation}
    \frac{\partial\mathcal{L}}{\partial W_{i+1}} = X_{i}^{T}\frac{\partial\mathcal{L}}{\partial Z_{i+1}},
    \label{eq:bp2}
\end{equation}
\begin{equation}
    \frac{\partial\mathcal{L}}{\partial X_{i}} = W_{i+1}^{T}\frac{\partial\mathcal{L}}{\partial Z_{i+1}},
    \label{eq:bp3}
\end{equation}
\begin{equation}
    \frac{\partial\mathcal{L}}{\partial B_{i+1}} = f'(Z_{i+1})\frac{\partial\mathcal{L}}{\partial X_{i+1}},
    \label{eq:bp4}
\end{equation}
Equations \ref{eq:bp1}-\ref{eq:bp4} allow the manipulation of the $i$-th layer gradients of the ANN from layer $i+1$, according to a choice of gradient-descent optimizer. This means that the weights $W_k$ and biases $B_k$ of the $k$-th training iteration are then updated to new values $W_{k+1}$ and $B_{k+1}$ for iteration $k+1$ such that they descend the gradient of $\mathcal{L}$. \texttt{refann} uses the Adam optimizer \cite{kingma2017adam}, and in this work, we find 30,000 training iterations are sufficient to minimize $\mathcal{L}$ over a given set of either LISA or ET $D_L$ and $\sigma_{D_L}$ data.

\begin{figure}[htb!]
    \centering
    \includegraphics[width=0.6\columnwidth]{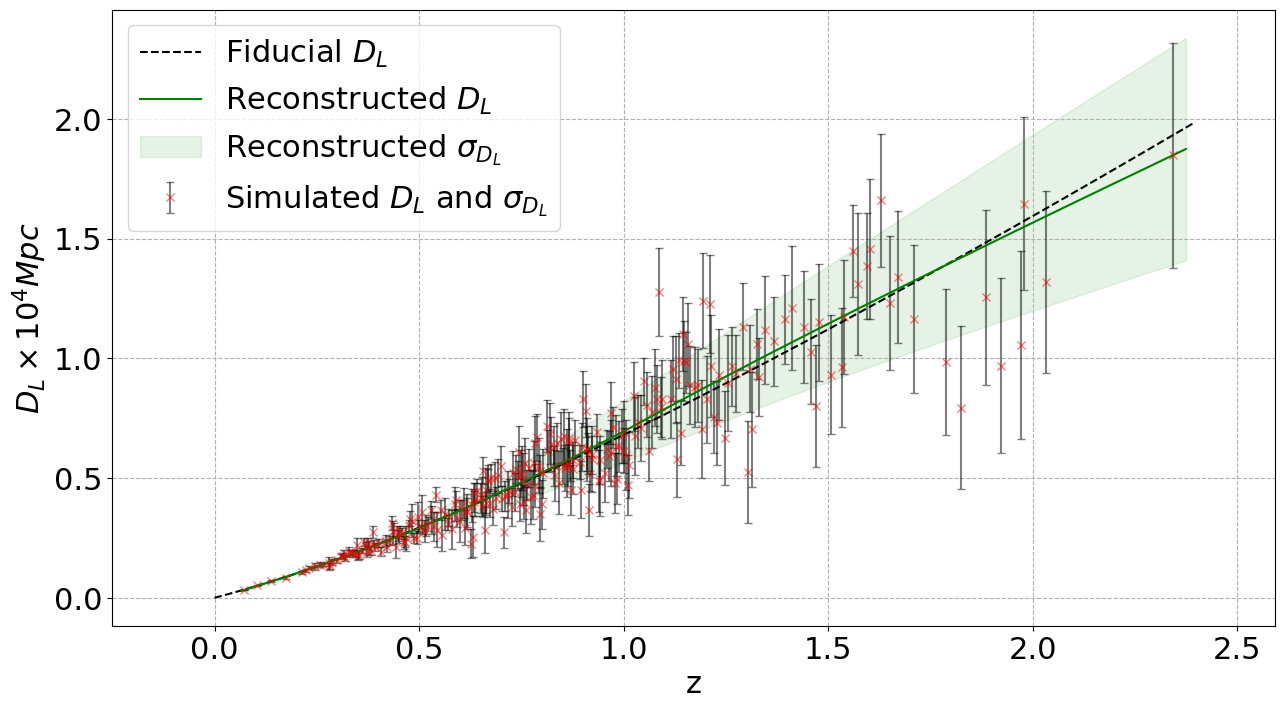}
    \includegraphics[width=0.6\columnwidth]{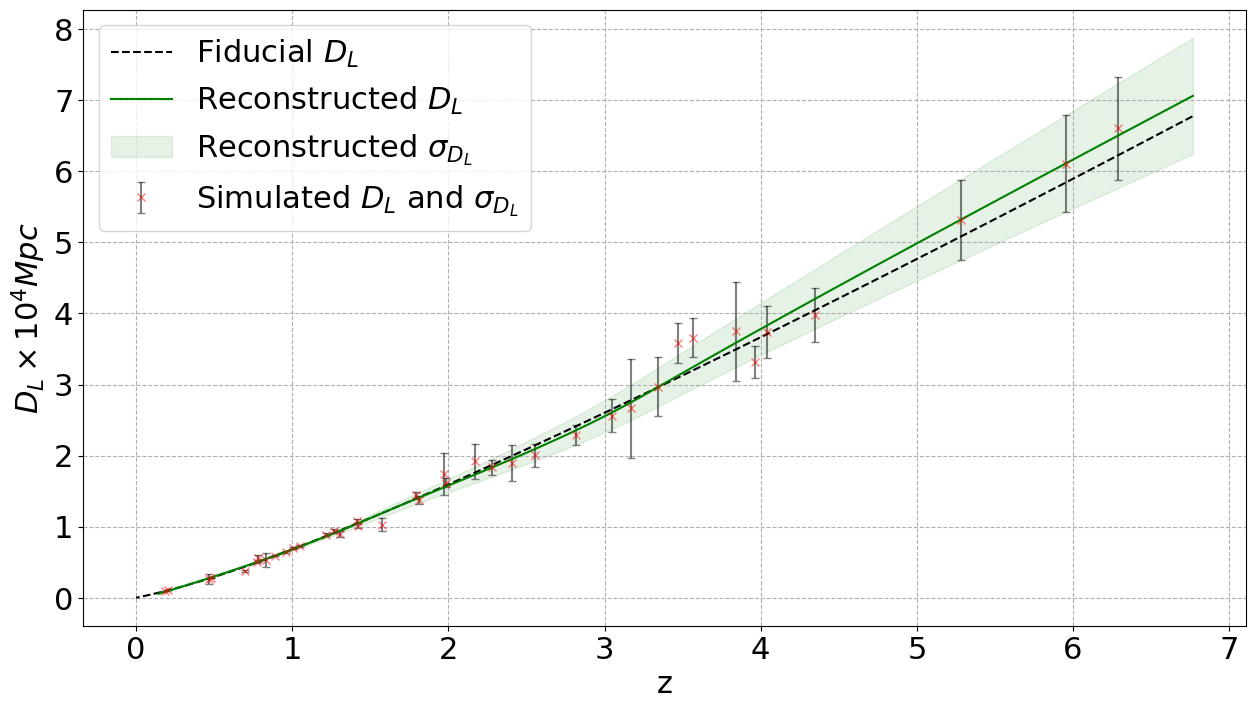}
    \caption{Reconstruction of $D_{\rm L}(z)$ and $\sigma_{D_L}(z)$ from simulated ET gravitational wave sirens in the range $0 \leq z \leq 2.5$ (top panel) and from simulated LISA gravitational wave sirens in the range $0 \leq z \leq 8$ (bottom panel).}
    \label{fig:GW_recon}
\end{figure}

Fig.~\ref{fig:GW_recon} shows the functions $D^{\rm ANN}_{\rm L}(z)$ and associated errors $\sigma^{\rm ANN}_{D_L}(z)$ given by models trained on ET and LISA data, respectively. Overall, ET is expected to observe a larger number of GW events from which we can extract a greater number of $D_{\rm L}(z)$ measurements compared to LISA. However, it is worth noting that on average, ET has higher uncertainties $\sigma_{D_L}(z)$ at lower redshifts. In contrast, LISA provides more accurate $D_{\rm L}(z)$ measurements across a larger redshift range. Regarding the reconstruction, the ET model fits the majority of the data, although some extraneous data points evade capture. Conversely, the LISA model captures the simulated observational data well but displays significantly higher uncertainties at $z > 6$ due to the sparsity of events in this redshift range.

\subsection{$r_d$ from $D^{\rm ANN}_{\rm L}(z)$ and $\theta_{BAO}(z)$}
Using the ANN model trained on either LISA or ET data, we can reconstruct a continuous function, $D_{\rm L}^{\rm ANN}(z)$, that can be evaluated at the forecasted $\theta_{\rm BAO}(z)$ redshift values. Therefore, we can combine these data to obtain a model-independent estimate of the sound horizon at the baryon-drag epoch, $r_d$, and its uncertainty, $\sigma_{r_{\rm d}}$:
\begin{equation}
r_{\rm d} = \frac{\theta_{\rm BAO}(z)D_{\rm L}^{\rm ANN}(z)}{(1+z)}, 
\end{equation}
\begin{equation}
\sigma_{r_{\rm d}} = \frac{\sqrt{(D_{\rm L}^{\rm ANN}(z) \sigma_{\theta_{\rm BAO}})^2 + (\theta_{\rm BAO}(z) \sigma_{D_{\rm L}^{\rm ANN}})^2}}{(1+z)}
\end{equation}

\begin{figure*}[htb!]
    \centering
    \includegraphics[width=\textwidth]{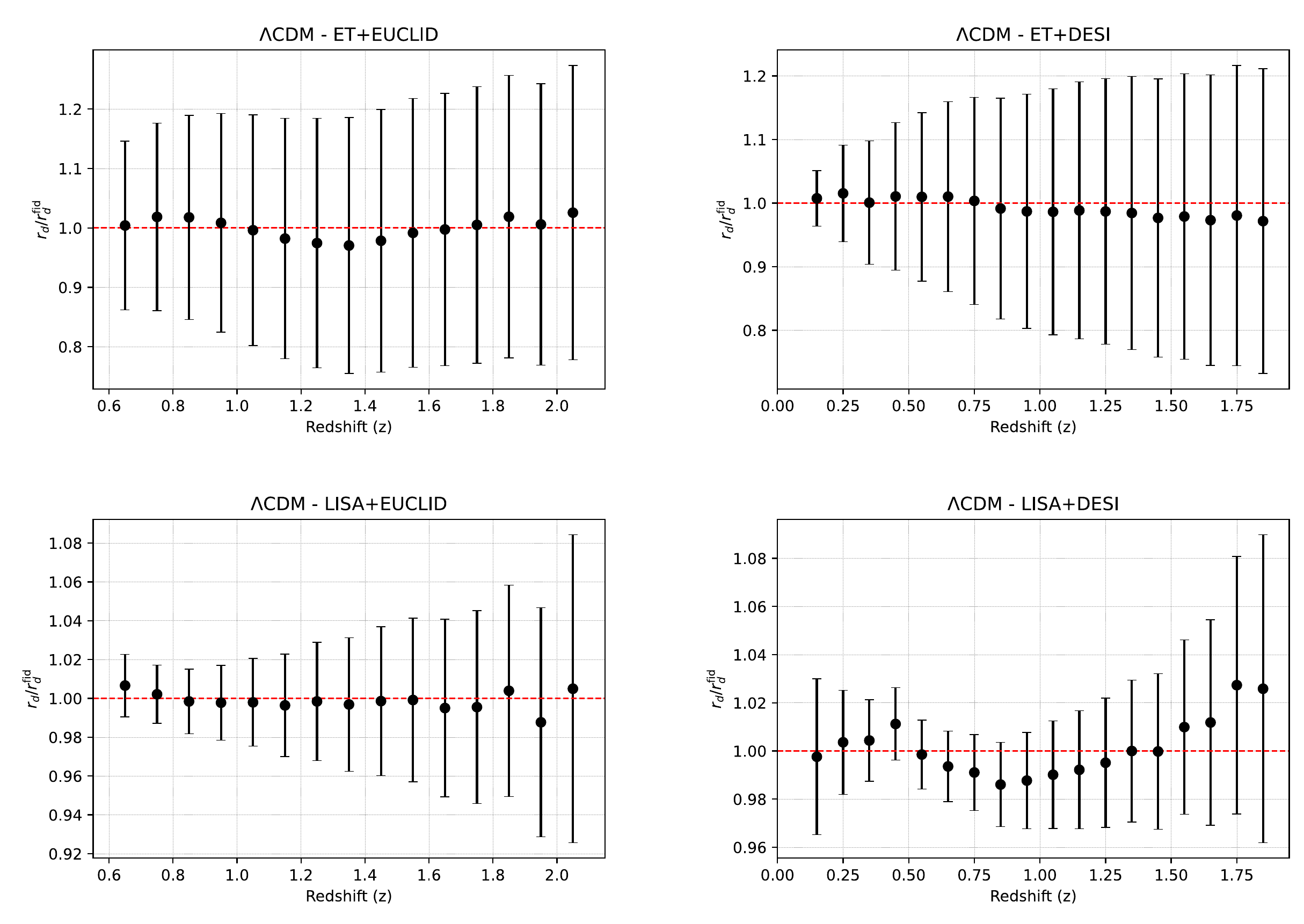}
    \caption{Estimates of $r_d / r^{\rm fid}_d$ for the four different pairs of datasets resulting from forecasts of gravitational wave experiments (ET and LISA) and baryon acoustic oscillation surveys (DESI and Euclid).}
    \label{fig:rd_recon}
\end{figure*}

Using forecasts from gravitational wave experiments (ET and LISA) along with forecasts for the angular scale of baryon acoustic oscillations (DESI and Euclid), we have four different pairs of datasets:
\begin{itemize}
    \item $D_{\rm L}^{\rm ANN}(z)$ from ET data and $\theta_{\rm BAO}(z)$ from DESI data.
    \item $D_{\rm L}^{\rm ANN}(z)$ from ET data and $\theta_{\rm BAO}(z)$ from Euclid data.
    \item $D_{\rm L}^{\rm ANN}(z)$ from LISA data and $\theta_{\rm BAO}(z)$ from DESI data.
    \item $D_{\rm L}^{\rm ANN}(z)$ from LISA data and $\theta_{\rm BAO}(z)$ from Euclid data.
\end{itemize}
For any of these four combinations of data, we can obtain several independent estimates of $r_{\rm d}$, along with its corresponding uncertainty. Specifically, we can obtain an estimate for each BAO data point. In the manuscript (Tab.1), we provide the relative precision $\sigma_{r_{\rm d}}/r_{\rm d}$ that we can achieve at different redshifts $z$. Instead, in Fig.~\ref{fig:rd_recon}, we show realistic estimates of $r_d / r^{\rm fid}_d$ for the four different pairs of datasets described above.

\clearpage

\section{Consistency Tests and comparison with the Literature}
\label{sec:tests}
\subsection{Constraining the Hubble Parameter}
Our method of forecasting the angular BAO scale and the redshift-luminosity distance relation $D_L(z)$ from gravitational wave events does introduce some artificial randomness in the Gaussian noise used to simulate realistic mock data. As explained in the manuscript, we considered different statistical metrics to introduce a controlled --yet realistic-- scatter of data points around the fiducial $\Lambda$CDM cosmology. We explicitly test whether the random noise introduced is indeed under control by performing an MCMC test with forecasted data, aimed at checking whether we can recover the fiducial cosmology afterward.

In particular, the simulated BAO and GW data are compared to the theoretical $\theta(z)$ and $D_L(z)$ values through a multivariate Gaussian likelihood:
\begin{equation}
    -2\ln \mathcal{L} = \sum (\mathbf{\mu} - \hat{\mathbf{\mu}})C^{-1}(\mathbf{\mu} - \hat{\mathbf{\mu}})^T ~,
\end{equation}
where $\mathbf{\mu}$ and $\hat{\mathbf{\mu}}$ are the vectors containing the simulated and theoretical values $\theta(z)$ and $D_L(z)$ at each redshift, and $C$ is their simulated covariance matrix.\footnote{In our mock data, we adopt a diagonal covariance matrix.}

For the different GW and BAO surveys, we implement the respective likelihood functions in \texttt{cobaya}~\cite{Torrado_2021}\footnote{\url{https://github.com/CobayaSampler/cobaya}} and perform a Background-only MCMC analysis. Specifically, we leave $H_0$ and $\Omega_c h^2$ free to vary within flat ranges $H_0\in[20,100]$ km/s/Mpc and $\Omega_c h^2\in[0.001,0.99]$. Instead, we assume a BBN prior $\Omega_b h^2=\mathcal{G}(0.0224,0.0001)$, and fix the following parameters to their best-fit values: $\log(10^{10}A_s)=3.0447$, $n_s=0.9659$, and $\tau=0.0543$.

\begin{figure*}[tb]
    \centering
    \includegraphics[width=\textwidth]{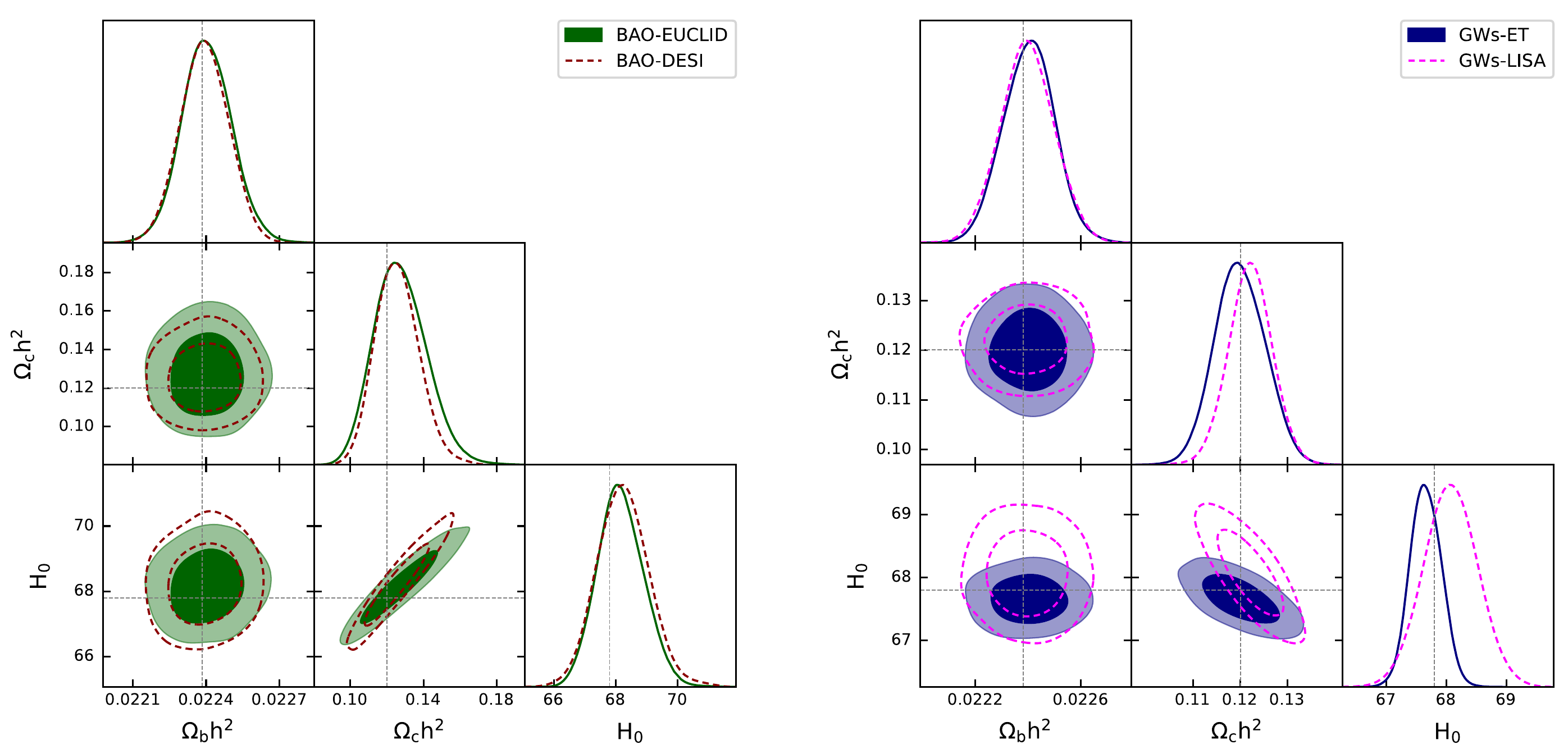}
    \caption{\textit{Left panel:} 1D posterior distribution functions and the 2D marginalized probability contours for the parameters of interest in our tests, obtained through MCMC analysis of the DESI and Euclid BAO mock data. \textit{Right panel:} 1D posterior distribution functions and the 2D marginalized probability contours for the same parameters obtained through MCMC analysis of the ET and LISA GW mock data.}
    \label{fig:MCMC}
\end{figure*}

The 1D posterior distribution functions and the 2D marginalized probability contours for the parameters of interest in our tests, obtained through MCMC analysis, are shown in Fig.~\ref{fig:MCMC}. As is clear from the figures, the fiducial values are always recovered well within the 68\% confidence level interval.

\subsection{Constraining power of GWs across different redshifts}
\label{sec.GWs_tests}

As we already pointed out, we are working within conservative assumptions, in a setting that is somewhat more cautious than what is typically assumed in the literature.\footnote{Note that we compared our results with existing literature and cross-checked our GW sky localization assumptions. Our forecasted localization accuracy remains equally or more conservative than many results obtained using state-of-the-art codes in the field, such as Bayestar (\url{https://lscsoft.docs.ligo.org/ligo.skymap/bayestar/index.html}).} However, to further validate our approach, we performed several analyses to identify the part of the GW dataset that is most important for our methodology, and  explicitly check that our approach accounts for all uncertainties and caveats surrounding assumptions on the population and merger rates of these massive binaries, as well as assumptions regarding astrophysical modeling and the ability to detect an EM counterpart. 

To strengthen the robustness of our analysis, we want to prove three points:
  
\begin{enumerate}

\item The GW detections that are most important for our consistency test to work are those at very low redshift ($z<0.5$). This is a crucial aspect because detecting an EM counterpart is expected to be significantly more feasible at lower redshifts than at higher redshifts.

\item Even assuming a very conservative number of 7-10 detected GW events at $z<0.5$ from LISA (disregarding all GW events at $z>0.5$), our methodology remains powerful enough to yield conclusive answers about the value of the sound horizon.

\item In our full analysis, we remained conservative in incorporating all limitations and caveats surrounding high-redshift GW detections. In particular, disregarding low-redshift GWs (i.e., detections at $z<0.5$), the combination of future surveys leads to uncertainties on $r_d$ that can be as large as 5-10\%, mostly driven by uncertainties on the value of $H_0$ inferred from GWs (thus recovering the results documented in the literature).

\end{enumerate}

\subsubsection{LISA constraining power across different redshifts}

\begin{figure*}[ht!]
    \centering
    \includegraphics[width=\textwidth]{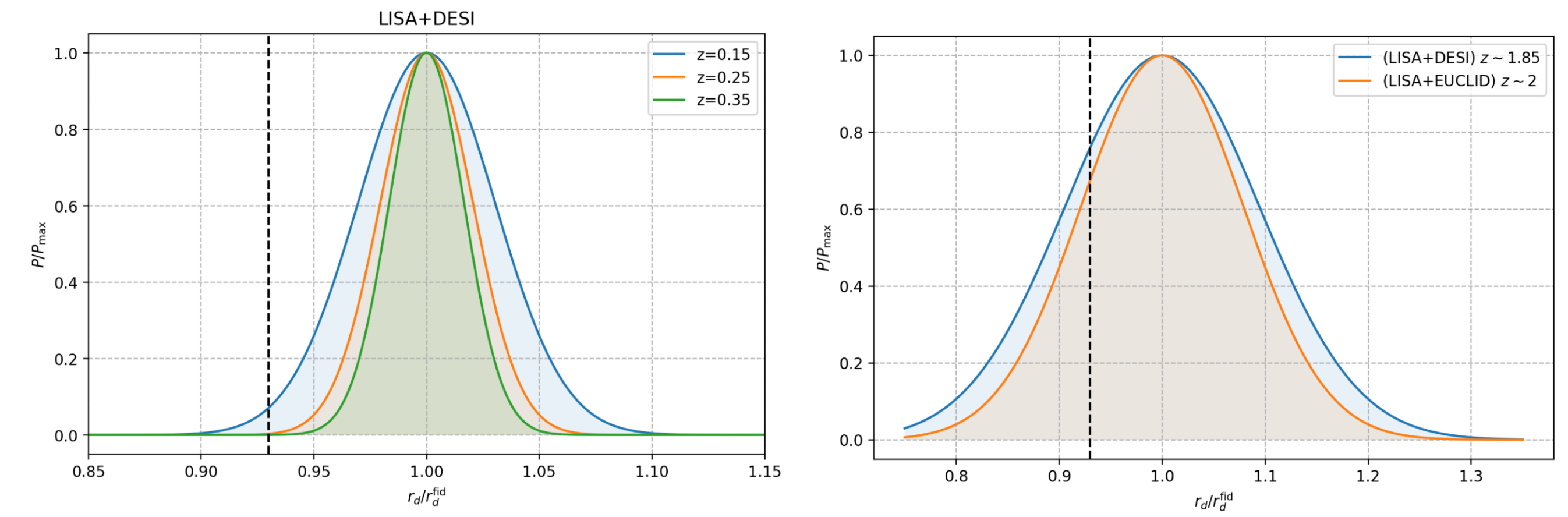}
    \caption{\textit{Left panel:} Forecasted 1D probability distribution function for $r_d/r_d^{\rm fid}$ obtained from the joint analysis of mock data from LISA and DESI experiments, retaining information up to $z=0.15$ (blue), $z=0.25$ (orange), and $z=0.35$ (green). For LISA, we restricted the analysis to only seven events with an EM counterpart at $z < 0.5$, which were used to reconstruct $D_L(z)$ up to the three redshifts outlined above. We disregarded all forecasted gravitational wave observations at $z > 0.5$, where detecting an EM counterpart is more challenging. While this slightly increases the uncertainties on $r_d/r_d^{\rm fid}$, low-redshift data are sufficient for applying our methodologies and can either confirm or rule out a reduction in the sound horizon needed to address the Hubble tension (black dashed line). \textit{Right panel:} Forecasted 1D probability distribution function for $r_d/r_d^{\rm fid}$ obtained from the joint analysis of mock data from LISA and DESI (blue) or EUCLID (orange) experiments, retaining information at $z \gtrsim 1$. For LISA, we restricted the analysis to detected events at $z > 1$, which were used to reconstruct $D_L(z)$. We disregarded all forecasted gravitational wave observations at $z < 0.5$. Focusing only on high-redshift data significantly increases the uncertainties on $r_d/r_d^{\rm fid}$ up to $\sim 10\%$, primarily due to the more challenging measurement of $H_0$ required to anchor BAO, which measures the combination $r_d \times h$. }
    \label{fig:test_1}
\end{figure*}

To prove the first two points (i.e., Points 1 and 2), we performed an additional robustness test by adopting conservative assumptions, considering only the 7 GW events forecasted for LISA at $z < 0.5$, where EM counterparts are more easily identified. Specifically, we reconstructed the luminosity distance using all GW observations in the range $0 < z < 0.5$, explicitly excluding any GW detections at higher redshifts. These 7 GW measurements provide a sufficiently precise luminosity distance reconstruction at $z < 0.5$, serving as a robust anchor for BAO data from DESI-like experiments. Our results are summarized in the left-side panel of Fig.~\ref{fig:test_1}. For the combined GW+BAO analysis, we incorporated the reconstructed luminosity distance up to three pivot redshifts ($z < 0.15$, $z < 0.25$, and $z < 0.35$) to forecast constraints on $r_d/r_d^{\rm fid}$. Our results show that the uncertainty on $r_d$ at $z = 0.35$ increased only slightly from $1.4\%$ to $1.6\%$. However, as illustrated in the left-side panel of Fig.~\ref{fig:test_1}, this level of precision remains sufficient to distinguish between models of new physics proposed to resolve the Hubble tension. These models predict a $\sim 7\%$ reduction in the sound horizon (dashed black line in the left-side panel of Fig.~\ref{fig:test_1}) and can still be ruled out (or confirmed) at more than $4\sigma$ significance.

To further corroborate Point 1 and to prove Point 3, we repeated a similar analysis to highlight the limitations of high-redshift standard sirens and argue that, beyond $z\sim1$, the precision on $r_d$ constraints is fundamentally limited to the $5-10\%$ level, in line with what is found for $H_0$ in the literature.\footnote{Note that our results already reflect this limitation: as shown in Table I of the main paper, the relative error on $r_d$ significantly increases at higher redshifts. Specifically, for a combination of LISA and DESI at $z\sim1.85$, we find a relative precision of approximately $6\%$, largely relaxed due to the weaker constraints on $H_0$ from LISA-like experiments at those redshifts. Since BAO experiments measure the combination $r_d \times h$, the constraint on $r_d$ is inherently tied to the precision of the GW-derived $H_0$ measurement, which degrades at high redshift. Similarly, for the combination of LISA and Euclid at $z\sim2$, the precision on $r_d$ remains within the $5-10\%$ range, again due to the combined uncertainties on $H_0$ from LISA and $r_d \times h$ from Euclid-like BAO measurements. This demonstrates that our results are in line with the limitations outlined in the literature, see e.g., \cite{Speri:2020hwc,Mangiagli:2023ize}.}

To illustrate this point, we conducted an additional analysis analogous to the one performed previously. Namely, we re-ran our data analysis pipeline, this time removing all LISA data at $z<1$, to assess the impact of retaining only high-redshift data ($z\gtrsim1$). We then reanalyzed the combination of LISA with both DESI-like and Euclid-like experiments to ensure that our results align well with those discussed in the literature. The results are summarized in the right-side panel of Fig.~\ref{fig:test_1}. For both cases, we find that the relative error on the sound horizon remains at the $\sim10\%$ level, which is fully consistent with the uncertainties on $H_0$ from LISA as reported in the literature, see e.g.,~\cite{Speri:2020hwc,Mangiagli:2023ize}. 

These result further highlights a key aspect of our methodology: the constraining power on $r_d$ is significantly driven by low-redshift GW-BAO cross-calibration, where standard sirens are expected to be more reliable due to a higher likelihood of detecting EM counterparts. If we rely only on high-redshift standard sirens, we would no longer be able to apply our methodology effectively, precisely because our modeling of the GW population already accounts for these limitations (and in part because BAO data are also expected to lose sensitivity at high redshift). As a result, when restricting the analysis to $z>1$, our pipeline naturally recovers a level of constraining power comparable to what has been documented in the literature, reinforcing the robustness of our approach.

\subsubsection{ET constraining power across different redshifts}
To further validate Point 1 -- namely, that GW observations at $z < 0.5$ play the most significant role in our methodology -- we perform a final check for ET, analogous to the one carried out for LISA. Compared to LISA, ET has a poorer angular resolution, making the identification of electromagnetic counterparts at high redshift significantly more challenging. We therefore repeat the same analysis using forecasted data for ET and focus on a conservative scenario in which only GW events with EM counterparts at low redshift are retained, explicitly removing all detections at $z > 0.5$.

We include only those events predicted to be observed at $z < 0.5$, where the association with an EM counterpart is more plausible. Using these low-redshift events, we reconstruct the luminosity distance $D_L(z)$ and combine the resulting information with mock BAO data from DESI-like surveys. As in the LISA case, we consider three separate setups for the joint GW+BAO analysis, using the reconstructed luminosity distance up to $z < 0.15$, $z < 0.25$, and $z < 0.35$, respectively, to forecast constraints on $r_d/r_d^{\rm fid}$.

\begin{figure*}[ht!]
    \centering
    \includegraphics[width=0.7 \textwidth]{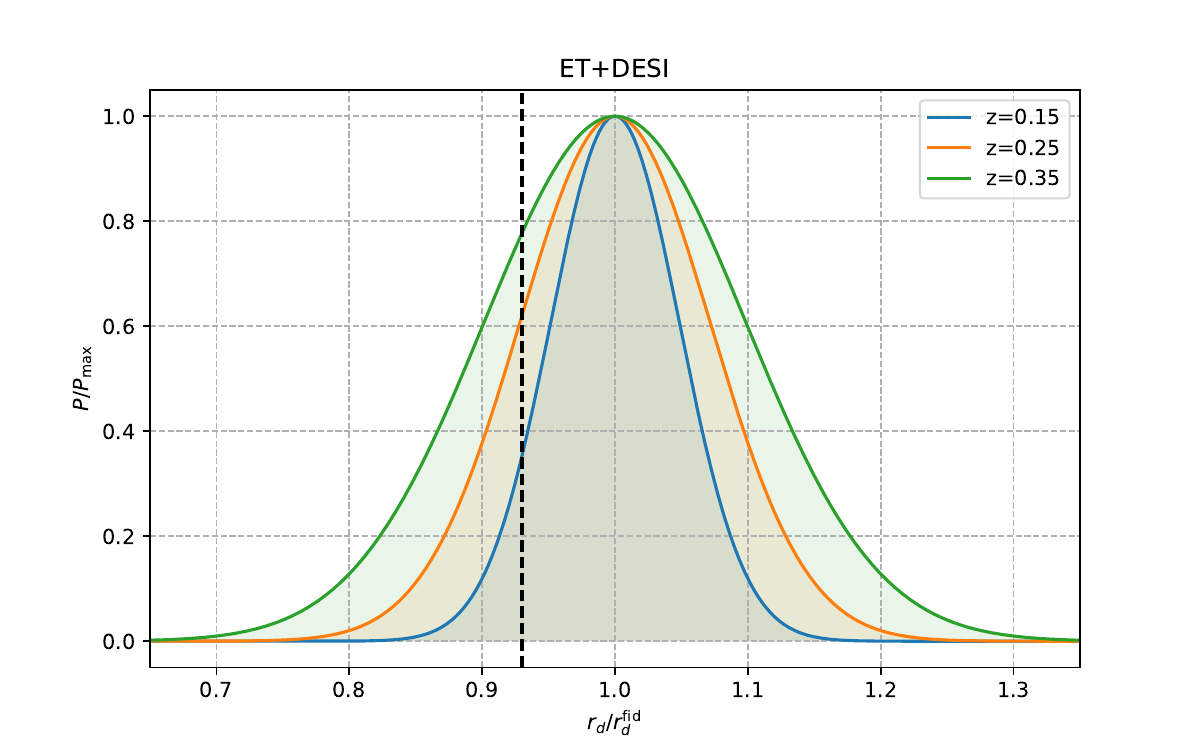}
    \caption{Forecasted 1D probability distribution function for $r_d/r_d^{\rm fid}$ obtained from the joint analysis of mock data from ET and DESI experiments, retaining information up to $z=0.15$ (blue), $z=0.25$ (orange), and $z=0.35$ (green). For ET, we restricted the analysis only to events with an EM counterpart at $z < 0.5$, which were used to reconstruct $D_L(z)$ up to the three redshifts outlined above. We disregarded all forecasted gravitational wave observations at $z > 0.5$, where detecting an EM counterpart is more challenging.}
    \label{fig:test_ET}
\end{figure*}

Our results are shown in the 1D probability distributions in Fig.~\ref{fig:test_ET}, which displays the posteriors on $r_d/r_d^{\rm fid}$ from the joint ET+DESI analysis. We find that the uncertainty on $r_d$ increases as higher redshift bins are included, reaching $4.8$\%, $7.2$\%, and $9.9$\% for the three cases above. Despite the mild increase in uncertainty, the results remain fully consistent with those reported in Table I of the main text for the ET+DESI combination (first three entries of the second column). This confirms that removing high-redshift events leads only to a moderate degradation in precision, and that the forecasts remain comparable to those obtained when including the full ET mock dataset, even with high-redshift detections.

That said, when focusing solely on ET, the constraining power on $r_d$ is likely not sufficient to confidently distinguish between models predicting a few-percent-level deviation from the Planck-$\Lambda$CDM sound horizon. Nonetheless, the combination of ET and DESI still enables a first model-independent measurement of $r_d$ at the $\sim 4$\% level, which already represents a valuable input for cosmological analyses.

In conclusion, this additional test confirms that, also in the case of ET, low-redshift GW events play a crucial role in anchoring the BAO scale to GW luminosity distance measurements, and are essential to fully exploit the cross-calibration strategy developed in this work.

\subsection{Dependence Modeling and Number of Multi-messenger Detections}
\label{sec.N_EM_detections}
In the previous subsection, we have shown that the low-redshift ($z < 0.5$) portion of the multi-messenger population plays a key role in anchoring our consistency test. This finding is particularly relevant, as the identification of EM counterparts is expected to be significantly more feasible at low redshifts, where localization and follow-up are observationally more accessible.

Given the central importance of this result, in this section we aim to reinforce our conclusions by examining how our methodology responds to variations in both the modeling and the number of available multi-messenger detections. To this end, we carry out an explicit stress test by deliberately considering pessimistic scenarios in which only a very limited number of GW events with EM counterparts are successfully identified at low redshift, and no detections are assumed at higher redshift. This allows us to push the method into a minimal-statistics regime and assess its robustness under extremely conservative (yet plausible) conditions.

Specifically, we fix the total number of GW events with EM counterparts detectable by LISA-like experiments, varying this number systematically from a maximum of 10 down to just 4. For each fixed number, we generate over 50 independent realizations of mock GW datasets using the simulation pipeline described in \autoref{section:appendix-GW}. These realizations incorporate stochastic variations in the redshift distribution of events, capturing different sampling densities, chance clustering at specific redshifts, and more uniform event spacing. This variability is especially important in the low-redshift regime and when the number of EM detections is small, as the precision on the luminosity distance $D_L$ (and consequently on $r_d$) can depend more significantly on how the events are distributed.\footnote{For example, a realization with detections clustered around a narrow redshift range may yield stronger constraints there but leave other intervals poorly constrained. Whether these clusters overlap with the redshifts of available BAO measurements can further improve or degrade constraints on the sound horizon scale. Conversely, more evenly spaced events tend to provide broader but shallower sensitivity across the full redshift range.} Additionally, other observational factors can also contribute to this variability. As shown in Fig.~\ref{fig:GWs}, the error bars on $D_L$ for low-redshift GW events are not uniform: favorable source orientations and lower redshifts can yield percent-level uncertainties, but precision can also vary due to source inclination, detector response patterns, and statistical fluctuations inherent to small-number samples. While these effects tend to average out when reconstructing the luminosity distance from the full mock dataset analyzed in the main text, they can have a non-negligible impact in the highly restrictive scenarios studied here.

\begin{figure*}[t!]
    \centering
    \includegraphics[width=0.8\textwidth]{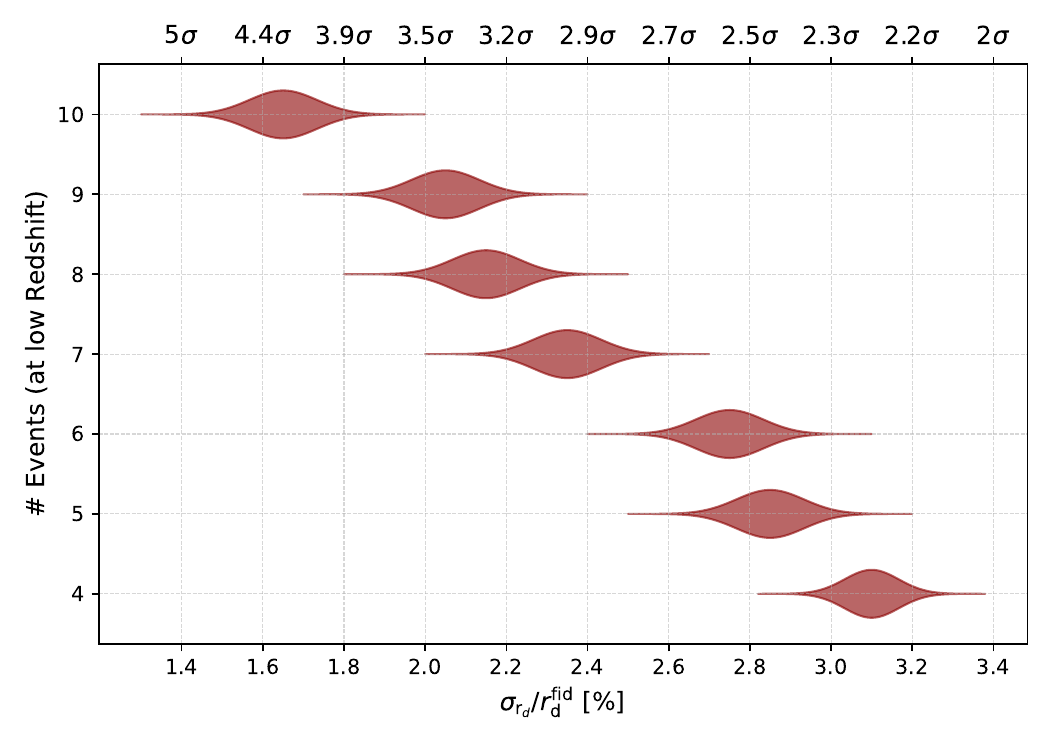}
    \caption{Forecasted relative precision on the sound horizon as a function of the number of low-redshift GW events with EM counterparts detected by a LISA-like experiment. Results are obtained by combining these GW events with DESI-like BAO measurements. For each fixed number of GW detections, we perform over 50 simulations, marginalizing over a range of stochastic factors, including the redshift distribution of the events (e.g., whether they are evenly spread or clustered by chance), source inclination, and detector-specific variability. Each case is shown as a probability distribution (dark red), reflecting the resulting spread in forecast precision. The top axis indicates the statistical significance at which cosmological models predicting a reduction of the sound horizon by up to 7\% could be tested or excluded.}
    \label{fig:test_2}
\end{figure*}

We then combine these 50 mock GW datasets with DESI-like BAO data to evaluate the consistency test under minimal and realistic detection conditions. For each fixed number of EM counterparts, we ultimately provide an interval distribution of the expected relative percentage precision on the sound horizon, effectively marginalizing over noise and sampling variance.

The results are shown in Figure~\ref{fig:test_2}, where each configuration is represented by a probability distribution for the relative precision on the sound horizon (shown in red). The bottom x-axis indicates the percentage relative precision achieved for each fixed number of low-redshift GW detections with an EM counterpart (listed on the y-axis). The top x-axis instead shows the corresponding statistical significance at which a model predicting a 7\% reduction in $r_d$ (consistent with what is typically required to resolve the Hubble tension) could be tested or excluded.

As evident from the Figure:

\begin{itemize}
\item Assuming 10 low-redshift detections our methodology performs remarkably well. Our simulations indicate that we can achieve a relative precision on the sound horizon of about 1.4\% up to approximately 2\%. This level of precision would allow us to distinguish models that shift the sound horizon by up to 7\% with high  4–5$\sigma$ significance comparable to the one obtained assuming high-redshift EM detections.
\item For 9 to 7 low-redshift EM detections, the method achieves a relative precision on $r_d$ ranging from 1.6\% to 2.6\%, corresponding to a 3-4$\sigma$ test of models relevant to addressing the Hubble tension. Thus the method remains highly informative under modest detection statistics.
\item In more pessimistic scenarios with only 6 to 4 EM counterparts, the uncertainty on $r_d$ increases in the range  between 2.4 and 3.5\%, allowing for a 2--3$\sigma$ level of discrimination. Despite this degradation, the methodology still enables a competitive measurement of the sound horizon, even under very pessimistic conditions.
\item Below 4 detections, our method becomes instead unreliable due to the difficulty of reconstructing the distance-redshift relation from such a limited number of events. In this case, alternative strategies could be considered, such as using the few available GW events to calibrate supernova data or including additional standardizable probes such as quasars and gamma-ray bursts. However, we consider these extended analysis beyond the scope of this work.
\end{itemize}

In light of these results, we conclude that while the number of observed events with an EM counterpart does affect the final precision, the methodology does not rely on a large number of detections. What is needed is \textit{enough} well-measured GW events to effectively anchor the BAO measurements. Our simulations show that 7 to 10 low-redshift detections would already provide ideal conditions. Even with 5 to 7 events, the method remains robust, enabling a model-independent test of new physics at greater than 3$\sigma$ significance. With 4 to 5 events, competitive precision is still achievable, potentially revealing interesting 2–3$\sigma$ hints of deviations from standard cosmology while for fewer than 4 events, additional standardizable probes would likely need to be incorporated for the method to remain informative. We note, however, that detecting $\gtrsim 4$ GW events with EM counterparts at $z \lesssim 0.5$ appears to be a highly achievable target for next-generation GW experiments.

\subsection{Independence on the Fiducial cosmology}
\label{sec.Independence_fiducial_cosmo}

Last but not least, we aim to clarify the dependence of our results on the choice of fiducial cosmology. The only aspect of our approach that requires an assumption about a fiducial cosmology is the simulation of data for future GW and BAO experiments. This step is necessary because generating mock observations requires an underlying cosmological model that provides theoretical predictions for the distances these experiments will measure.

GW and BAO experiments will provide measurements of cosmological distances, such as the angular scale of the sound horizon and the luminosity distance $D_L$ at different redshifts. The values of these quantities depend on the assumed cosmological model. Thus, when simulating the data these experiments will obtain, we must adopt a fiducial cosmology. However, this does not mean that our methodology itself is model dependent.

\begin{figure*}[t!]
    \centering
    \includegraphics[width=0.7\textwidth]{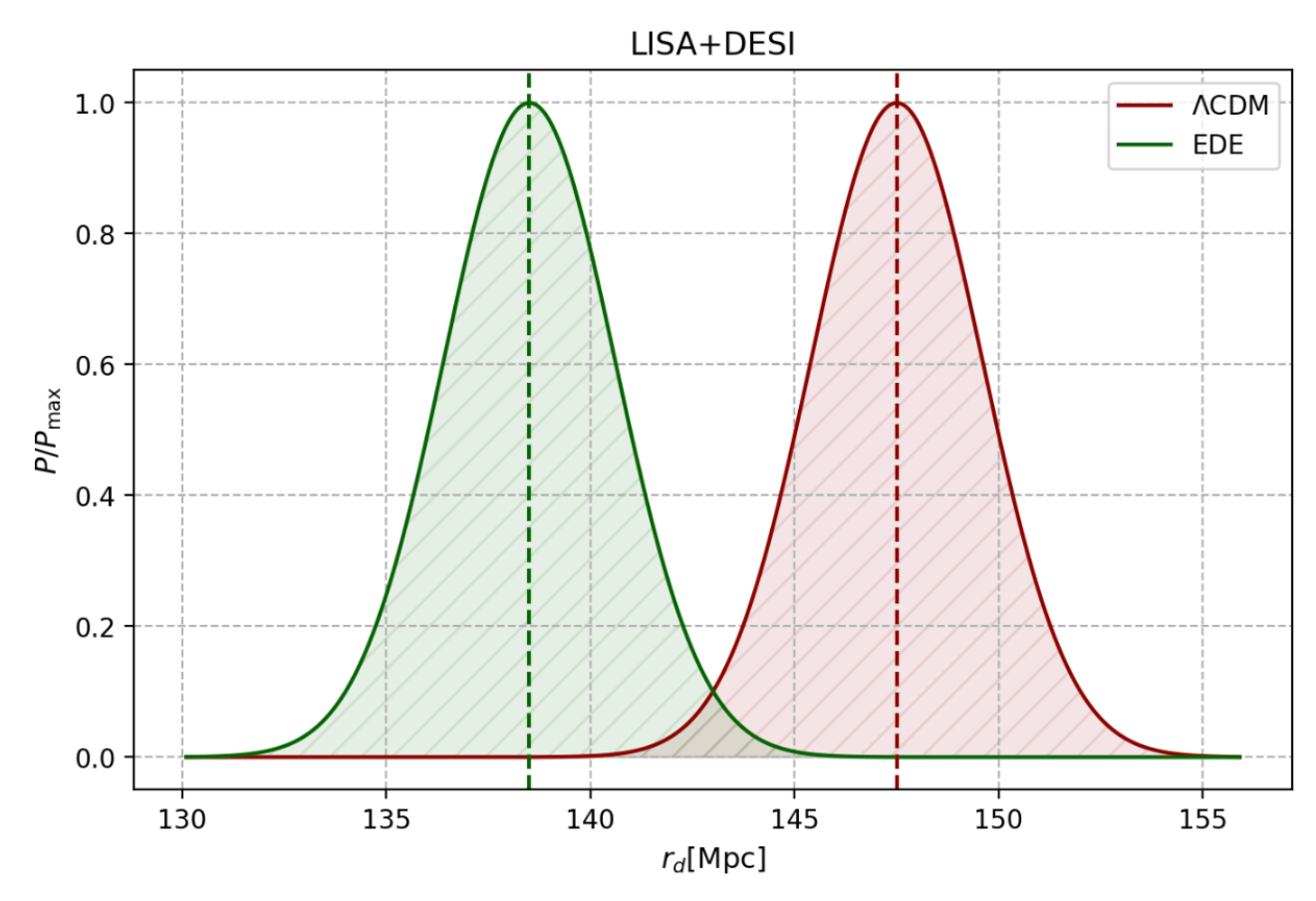}
    \caption{Forecasted 1D probability distribution function for $r_d/r_d^{\rm fid}$ obtained from the joint analysis of mock data from LISA and DESI experiments, retaining all the information as in our main analysis, and produced under two different fiducial cosmologies. We generated BAO and GW simulated data assuming a fiducial $\Lambda$CDM cosmology with a predicted sound horizon of $\sim 147$ Mpc (red dashed line). Given this set of simulated data, we applied our methodology to infer the sound horizon directly from the data, finding excellent agreement (red 1D posterior distribution). Similarly, we repeated the analysis by producing forecasted data for an Early Dark Energy universe, where the sound horizon is $r_d \sim 139$ Mpc. Again, our methodology successfully recovered this value (green 1D posterior distribution). Importantly, the precision with which we infer the value of the sound horizon (as well as our methodology) does not depend on the fiducial model assumed to generate the mock data.}
    \label{fig:test_3}
\end{figure*}

To demonstrate that our results are not contingent on the specific cosmological model assumed when generating mock data, we explicitly tested our methodology using different fiducial cosmologies. In our main analysis, we simulated BAO and GW data assuming $\Lambda$CDM with Planck best-fit parameters, where the fiducial sound horizon is $\sim 147$ Mpc (red dashed line in Fig.~\ref{fig:test_3}). Applying our pipeline, we successfully recovered this fiducial value with percent-level accuracy, as shown in the 1D posterior distribution function for $r_d$ in Fig.~\ref{fig:test_3}.

To further validate the robustness of our method, we repeated the analysis assuming an Early Dark Energy (EDE) fiducial cosmology, which introduces an additional phase of near de Sitter expansion before recombination. This model includes three extra free parameters compared to $\Lambda$CDM: the fraction of EDE, the initial displacement of the axion field driving the EDE phase, and the redshift at which the fraction of EDE reaches its maximum. We fixed these parameters such that the predicted value of $H_0 \sim 73$ km/s/Mpc, leading to a well-documented $\sim 7\%$ reduction in the sound horizon (green dashed line in Fig.~\ref{fig:test_3}). Simulating GW and BAO data under this EDE cosmology and applying our methodology, we found that we could successfully recover the fiducial sound horizon value with the same relative precision as in the $\Lambda$CDM case. 

This exercise demonstrates that our methodology remains valid and performs consistently regardless of the fiducial cosmology assumed in the simulations. Our approach is no more model dependent than standard techniques used in the GW standard siren community to extract $H_0$. In summary, future GW and BAO experiments will provide direct distance measurements at different redshifts without assuming a specific cosmological model. By combining these data, we will be able to measure the sound horizon and test whether the pre-recombination Universe follows $\Lambda$CDM or a model like EDE. The success and feasibility of our approach do not depend on the fiducial cosmology chosen for simulations, as we have explicitly demonstrated that the relative precision in extracting $r_d$ remains unchanged across different models.

\end{document}